\documentclass[10pt,twocolumn,letter]{article}

\usepackage{amssymb,amsmath}
\usepackage[in]{fullpage}
\usepackage{graphicx}
\usepackage{qcircuit}
\usepackage{color}
\usepackage{hyperref}

\DeclareMathOperator{\EE}{\mathbb{E}}

\newcommand{\set}[1]{\lbrace #1 \rbrace}
\newcommand{\abs}[1]{\lvert #1 \rvert}

\newcommand{\bra}[1]{\langle #1 |}
\newcommand{\ket}[1]{| #1 \rangle}
\newcommand{\innerprod}[2]{\langle #1 | #2 \rangle}

\newcommand{\norm}[1]{\lVert #1 \rVert}
\newcommand{\tensor}{\otimes}

\newcommand{\RR}{\mathbb{R}}
\newcommand{\CC}{\mathbb{C}}
\newcommand{\ZZ}{\mathbb{Z}}

\begin{document}

\title{An Uncertainty Principle for the Curvelet Transform, and the Infeasibility of Quantum Algorithms for Finding Short Lattice Vectors}

\author{Yi-Kai Liu$^{1,2}$\\
$^1$Applied and Computational Mathematics Division,\\
National Institute of Standards and Technology (NIST), Gaithersburg, MD, USA\\
$^2$Joint Center for Quantum Information and Computer Science (QuICS),\\
NIST/University of Maryland, College Park, MD, USA}

\date{\today}

\maketitle


\begin{abstract}
The curvelet transform is a special type of wavelet transform, which is useful for estimating the locations and orientations of waves propagating in Euclidean space. We prove an uncertainty principle that lower-bounds the variance of these estimates, for radial wave functions in $n$ dimensions. 

As an application of this uncertainty principle, we show the infeasibility of one approach to constructing quantum algorithms for solving lattice problems, such as the approximate shortest vector problem (approximate-SVP), and bounded distance decoding (BDD). This gives insight into the computational intractability of approximate-SVP, which plays an important role in algorithms for integer programming, and is used to construct post-quantum cryptosystems. 

In this approach to solving lattice problems, one prepares quantum superpositions of Gaussian-like wave functions centered at lattice points. A key step in this procedure requires finding the center of each Gaussian-like wave function, using the quantum curvelet transform. We show that, for any choice of the Gaussian-like wave function, the error in this step will be above the threshold required to solve BDD and approximate-SVP.
\end{abstract}


\section{Introduction}

Consider a function $f:\: \RR^n \rightarrow \RR$ that describes waves propagating in Euclidean space, in the sense that there exists a smooth $(n-1)$-dimensional surface $S$ in $\RR^n$, such that $f(x)$ (and its derivatives) are large when $x$ is near $S$. In physical terms, $S$ can be interpreted as the ``wavefront'' of a wave propagating in time. 

The curvelet transform is a type of wavelet transform that is useful for estimating properties of the set $S$ \cite{candes2005continuous}. We prove an uncertainty relation that sets an ultimate limit on the accuracy of these estimates, when $f$ is a radial function (see Section \ref{subsec-uncertainty}). 

This simple case is relevant to the study of quantum algorithms for solving lattice problems, such as the approximate shortest vector problem (approximate-SVP) and bounded distance decoding (BDD). These problems are conjectured to be hard, and they play an important role in algorithms for integer programming, and in a widely-studied class of post-quantum cryptosystems \cite{micciancio2002complexity, nguyen2010lll, peikert2016decade} (see Section \ref{subsec-qalgs-svp}). We consider a plausible approach to solving approximate-SVP and BDD on a quantum computer, using Gaussian-like lattice superposition states and the quantum curvelet transform. We then use our uncertainty relation to prove that this approach cannot succeed in solving approximate-SVP and BDD (see Section \ref{subsec-obstacles-solving-lattice-problems}).


\subsection{An Uncertainty Principle for the Curvelet Transform}
\label{subsec-uncertainty}

The curvelet transform works by decomposing $f$ into a sum of basis functions $\gamma_{a,b,\theta}$, called curvelets, where each curvelet is a wave-like function that is supported near a location $b \in \RR^n$, and oscillates in a direction given by a unit vector $\theta \in S^{n-1}$ (where $S^{n-1}$ denotes the unit sphere in $\RR^n$). In addition, each curvelet has a ``scale'' parameter $0<a\leq 1$, meaning that it is a supported on a plate-like region that has thickness $a$ in the direction $\theta$, and diameter $\sqrt{a}$ in the directions orthogonal to $\theta$. 

Formally, we define the continuous curvelet transform $\Gamma_f:\: (0,1) \times \RR^n \times S^{n-1} \rightarrow \CC$ by letting $\Gamma_f(a,b,\theta)$ be the coefficients in the decomposition of $f$ in the curvelet basis described above. By restricting $(a,b,\theta)$ to discrete domains in an appropriate way, we can also construct discrete curvelet transforms, which can be computed efficiently. 

The goal of this construction is to reveal information about the surface $S \subset \RR^n$ that describes the wavefront of $f$. The intuition is that $\Gamma_f(a,b,\theta)$ will be large when the location $b$ is close to $S$, the direction $\theta$ is orthogonal to $S$ (at the location $b$), and the scale $a$ matches the wavelength of the oscillations of $f$ at the location $b$. Furthermore, one expects that $b$ and $\theta$ will be more accurate when $a$ is small, i.e., when $f$ has high-frequency oscillations. This is reminiscent of Heisenberg's uncertainty principle, which states that if a function is mostly supported on a small region of space, then its Fourier transform must be spread out over a broad range of frequencies.

We prove an uncertainty relation for the curvelet transform, for the special case where $f$ is a radial function centered at some point $c \in \RR^n$, i.e., $f$ can be written in the form 
\begin{equation}
f(x) = f_0(\norm{x-c}_2), 
\end{equation}
where $f_0$ is some function $f_0:\: [0,\infty) \rightarrow \RR$, and $\norm{\cdot}_2$ denotes the $\ell_2$ norm on $\RR^n$. 
In this special case, the information revealed by the curvelet transform has a simple interpretation: we expect that $\Gamma_f(a,b,\theta)$ will be large when the location $b$ and the direction $\theta$ define a line 
\begin{equation}
\label{eqn-line}
\ell(b,\theta) = \set{b+\lambda\theta \;|\; \lambda \in \RR}
\end{equation}
that passes near the center point $c$. This gives us a simple way to formulate an uncertainty principle: we want to lower-bound the squared distance from the line $\ell(b,\theta)$ to the point $c$, which we denote $d(\ell(b,\theta),c)^2 = \min_{x\in\ell(b,\theta)} \norm{x-c}_2^2$. This can be written more simply as
\begin{equation}
d(\ell(b,\theta),c)^2 = (b-c)^T(I-\theta\theta^T)(b-c),
\end{equation}
which is the squared length of the vector $b-c$ projected onto the subspace orthogonal to $\theta$.

Formally, we show that there exist constants $C>0$ and $C'>0$ such that, for any radial function $f$ centered at $c \in \RR^n$, and for any $0<a\leq 1$, 
\begin{equation}
\label{eqn-curvelet-uncertainty}
\begin{split}
\int_{S^{n-1}} &\int_{\RR^n} (b-c)^T(I-\theta\theta^T)(b-c) \abs{\Gamma_f(a,b,\theta)}^2 db d\sigma(\theta) \\
&\geq Ca \int_{\RR^n} \norm{x-c}_2^2 \abs{\Xi_f(a,x)}^2 dx \\
&+ C'\lambda^2 n^2 a^2 \int_{S^{n-1}} \int_{\RR^n} \abs{\Gamma_f(a,b,\theta)}^2 db d\sigma(\theta).
\end{split}
\end{equation}
Here, $\Xi_f(a,\cdot)$ is defined in eq.~(\ref{eqn-xi-f}), and can be interpreted as the part of the function $f$ that lies within a particular scale $a$. The quantity $\lambda$ comes from the definition of the curvelet transform (see eq.~(\ref{eqn-chi})). It describes the frequency bands on which the curvelets are supported: for all $(a,b,\theta)$, the corresponding curvelet's Fourier transform $\hat{\gamma}_{a,b,\theta}$ is supported on an annulus $\set{k \in \RR^n \;|\; 1/e \leq \lambda a\norm{k}_2 \leq 1}$.

Eq.~(\ref{eqn-curvelet-uncertainty}) can be interpreted as a lower-bound on the uncertainty of a quantum measurement involving the curvelet transform, in the following way. Suppose that $f$ is normalized so that its $L_2$ norm is equal to 1. Then $f$ can be interpreted as a quantum wavefunction $\ket{f}$, which is a superposition over different values of $x$. Standard results about the curvelet transform (see eq.~(\ref{eqn-curvelet-plancherel})) imply that $\Gamma_f$ will also have $L_2$ norm equal to 1 (when parametrized appropriately). Hence $\Gamma_f$ can also be interpreted as a wavefunction $\ket{\Gamma_f}$, which is a superposition over different values of $(a,b,\theta)$. In a similar manner (see eq.~(\ref{eqn-xi-f-plancherel})), $\Xi_f$ can also be interpreted as a wavefunction $\ket{\Xi_f}$, which is a superposition over different values of $(a,x)$.

Now suppose we are given one copy of the quantum state $\ket{\Gamma_f}$, and we measure it, to obtain random measurement outcomes $a$, $b$ and $\theta$, which are modeled as classical random variables. Let $\EE_{\Gamma_f}$ denote an average over these random variables. In a similar manner, let $\EE_{\Xi_f}$ denote an average over random measurement outcomes obtained by measuring one copy of the quantum state $\ket{\Xi_f}$. Then eq.~(\ref{eqn-curvelet-uncertainty}) implies that, for all $0<\eta\leq 1$,
\begin{multline}
\label{eqn-curvelet-uncertainty-2}
\EE_{\Gamma_f}((b-c)^T(I-\theta\theta^T)(b-c) \;|\; a\geq\eta) \\
\geq C\eta \EE_{\Xi_f}(\norm{x-c}_2^2 \;|\; a\geq\eta) + C'\lambda^2 n^2\eta^2.
\end{multline}
(This follows by computing conditional probabilities, and using eq.~(\ref{eqn-xi-f-plancherel-a}).) This lower-bounds the expectation value of the squared distance $d(\ell(b,\theta),c)^2$ from the line $\ell(b,\theta)$ to the point $c$. This squared distance becomes smaller, when $\eta$ is smaller; but it is lower-bounded by a function of $\eta$, which is a sum of a linear term and a quadratic term. 

Eq.~(\ref{eqn-curvelet-uncertainty-2}) follows from a tighter, simpler analysis of the curvelet transform, building on techniques used in \cite{liu2009quantum}. However, note that this paper shows lower-bounds on the uncertainty, whereas \cite{liu2009quantum} showed upper-bounds. 

How strong is the lower-bound in eq.~(\ref{eqn-curvelet-uncertainty-2})? To answer this question, we also prove the following upper-bound: there exists a family of curvelets $\gamma_{a,b,\theta}$, and there exists a constant $C'>0$, such that for any radial function $f$ centered at $c \in \RR^n$, and for all $0<\eta\leq 1$,
\begin{multline}
\label{eqn-curvelet-upper-bound}
\EE_{\Gamma_f}((b-c)^T(I-\theta\theta^T)(b-c) \;|\; a\leq\eta) \\
\leq \tfrac{1}{\sqrt{2}}\eta \EE_{\Xi_f}(\norm{x-c}_2^2 \;|\; a\leq\eta) + C'\lambda^2 n^2\eta.
\end{multline}
This can be viewed as a generalization of the upper-bounds shown previously in \cite{liu2009quantum}. 

This shows that our uncertainty principle in eq.~(\ref{eqn-curvelet-uncertainty-2}) is optimal, up to a factor of $\eta$. In particular, eq.~(\ref{eqn-curvelet-uncertainty-2}) has the correct dependence on the dimension $n$, which is our main concern in this paper (and in \cite{liu2009quantum}).


\subsection{Quantum Algorithms for Lattice Problems}
\label{subsec-qalgs-svp}

Given a set of basis vectors $b_1,\ldots,b_n \in \RR^n$, let $\mathcal{L}(b_1,\ldots,b_n)$ be the lattice consisting of all integer linear combinations of the vectors $b_1,\ldots,b_n$, that is, 
\begin{equation}
\mathcal{L}(b_1,\ldots,b_n) = \set{\sum_{j=1}^n x_j b_j \;|\; x_1,\ldots,x_n \in \ZZ} \subset \RR^n.
\end{equation}
In cases where we do not need to show the basis vectors explicitly, we will simply write $L$ to denote a lattice $L = \mathcal{L}(b_1,\ldots,b_n)$. 

We say that a quantity $q$ is ``non-negligible'' if there exists a polynomial $p$ such that, in the limit as $n$ grows large, $q \geq 1/p(n)$. We will also write $q \geq 1/\text{poly}(n)$ to describe this situation. We say that a quantity $q$ is ``negligible'' if for all polynomials $p$, in the limit as $n$ grows large, $q \leq 1/p(n)$.

For any $\gamma \geq 1$, the $\gamma$-approximate Shortest Vector Problem (SVP) is defined as follows: 
\begin{quote}
Given a basis $b_1,\ldots,b_n \in \RR^n$, find a vector $v$ in the lattice $L = \mathcal{L}(b_1,\ldots,b_n)$ such that $\norm{v}_2 \leq \gamma \lambda_1(L)$,
\end{quote}
where $\lambda_1(L)$ is the length of the shortest nonzero vector in the lattice $L$, using the $\ell_2$ norm $\norm{\cdot}_2$. We will consider the case where $\gamma$ grows polynomially with $n$ (and $\gamma \geq \Omega(\sqrt{n})$). 

$\gamma$-approximate SVP is related to several other problems involving lattices, in particular, Bounded Distance Decoding (BDD). The BDD problem involves a parameter $r \in [0,\tfrac{1}{2})$, called the decoding radius. To emphasize this, we will sometimes call the problem ``radius-$r$ BDD.'' The problem is defined as follows: 
\begin{quote}
Given a basis $b_1,\ldots,b_n \in \RR^n$, and a vector $t\in\RR^n$ that is promised to be ``close'' to the lattice $L = \mathcal{L}(b_1,\ldots,b_n)$, find the lattice point $x\in L$ that is closest to $t$.
\end{quote}
Here $t$ is called the ``target vector,'' and ``close to the lattice'' means that there exists a lattice point $x\in L$ such that $\norm{x-t}_2 \leq r\lambda_1(L)$, where $\lambda_1(L)$ is the length of the shortest nonzero vector in $L$. If one does not have the promise that $t$ is close to the lattice, this problem is called the Closest Vector Problem (CVP).

Radius-$r$ BDD (with $r$ scaling inverse-polynomially with $n$) is closely related to $\gamma$-approximate SVP (with $\gamma$ scaling polynomially with $n$) \cite{lyubashevsky2009bounded}. In particular, the existence of an efficient algorithm for one problem implies an efficient algorithm for the other problem. 

These problems are believed to be computationally intractable: in the regime where $1/r$ and $\gamma$ are constants independent of $n$, these problems are NP-hard \cite{khot2005hardness}; and even in the easier regime where $1/r$ and $\gamma$ scale polynomially with $n$, where these problems are unlikely to be NP-hard \cite{aharonov2005lattice}, no efficient algorithms are known for solving these problems. 

Despite their intractability, these problems have been studied intensely in integer programming and combinatorial optimization \cite{nguyen2010lll}. More recently, the apparent hardness of these problems has been used to construct lattice-based cryptosystems, for post-quantum cryptography, fully-homomorphic encryption, and classical verification of quantum computations \cite{peikert2016decade}. These constructions work in the regime where $1/r$ and $\gamma$ scale polynomially with $n$. In order to have confidence in the security of these cryptosystems, it is important to study the possible approaches to designing quantum algorithms for solving these problems, in order to understand the reasons why these approaches do not succeed. 

In this paper we will study one such approach to designing quantum algorithms for lattice problems, which is based on preparing and measuring quantum superpositions of Gaussian-like wave functions centered at lattice points. This approach has been studied before, and it has been used to prove upper-bounds on the computational complexity of lattice problems. 

First, this approach has been used to show that radius-$r$ BDD (with $r$ scaling inverse polynomially with $n$) is in the intersection of the complexity classes NP and co-QMA \cite{aharonov2003lattice}, and in fact, in the intersection of NP and co-NP \cite{aharonov2005lattice}, which implies that the problem is unlikely to be NP-hard. Second, this approach has been used to show a quantum \textit{reduction} from $\gamma$-approximate SVP to a problem called Learning with Errors (LWE), that is, one can show a polynomial-time quantum algorithm that makes use of an oracle for solving the LWE problem, in order to solve $\gamma$-approximate SVP \cite{regev2009lattices}. This can be interpreted as an upper-bound on the complexity of $\gamma$-approximate SVP, or as a lower-bound on the complexity of LWE.

This quantum reduction to LWE plays an important role in analyzing the security of lattice-based cryptosystems, because it is a worst-case-to-average-case reduction. Essentially, this reduction shows that breaking random instances of these cryptosystems is at least as hard (in the sense of computational intractability) as solving the hardest instances of $\gamma$-approximate SVP, on a quantum computer. 

Finally, quite recently, these techniques have been applied to show a reduction to a novel continuous variant of the LWE problem \cite{bruna2021continuous}, and to construct efficient quantum algorithms for solving BDD with subexponential approximation factors, in special classes of lattices having abelian group structure \cite{eldar2022efficient}.

But this line of work has not resulted in an efficient quantum algorithm for \textit{general} lattices; nor does it provide evidence \textit{against} the possibility of such an algorithm. Our contribution in this paper is to identify one obstacle that helps explain the apparent absence of such an algorithm.



\subsection{Obstacles to Solving Lattice Problems}
\label{subsec-obstacles-solving-lattice-problems}

We begin with an informal description of our results, while deferring certain technical details to later sections of this paper.

\begin{figure}
\centering
\includegraphics{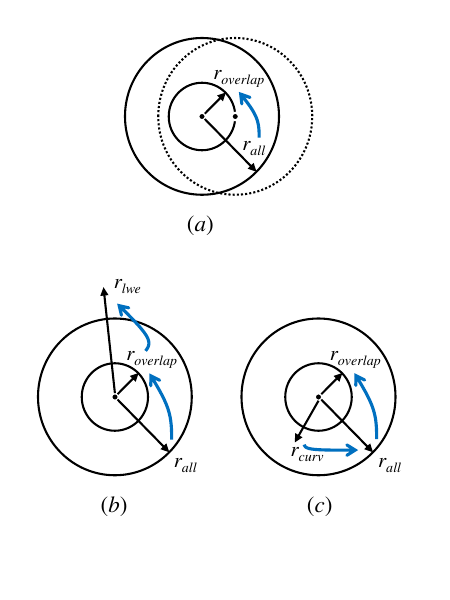}
\caption{Using Gaussian-like lattice superposition states $\ket{\psi}$ to solve the bounded distance decoding problem (BDD). Each panel shows the neighborhood surrounding a single lattice point. Without loss of generality, we can assume that $\lambda_1(L) = 1$. Panel $(a)$: $r_\text{overlap}$ is the largest distance such that, if $\ket{\phi}$ (dashed lines) is a copy of $\ket{\psi}$ shifted by distance $r_\text{overlap}$, then the overlap $\innerprod{\phi}{\psi}$ is non-negligible. $r_\text{all}$ is the smallest radius around the lattice that contains almost all of the probability amplitude in the state $\ket{\psi}$. Given an oracle that solves BDD with decoding radius $r_\text{all}$, one can prepare the state $\ket{\psi}$, and then solve BDD with decoding radius $r_\text{overlap}$ (thick blue arrow). But this is not useful, because $r_\text{overlap} < r_\text{all}$. Panel $(b)$: Using an LWE oracle \cite{regev2009lattices}, one can solve BDD with a larger decoding radius $r_\text{lwe} > r_\text{all}$ (thick blue arrow). By iterating this procedure, one can solve harder and harder instances of BDD, with larger and larger decoding radii. Panel $(c)$: Alternatively, by using the quantum curvelet transform \cite{liu2009quantum}, one can use a weaker oracle that solves BDD with decoding radius $r_\text{curv} < r_\text{all}$, to prepare the state $\ket{\psi}$ (thick blue arrow). If $r_\text{curv} < r_\text{overlap}$, then this would lead to a quantum algorithm for solving BDD. In this paper, we prove that this is impossible, for a large class of Gaussian-like lattice superposition states $\ket{\psi}$.}
\label{fig-radii}
\end{figure}

First, we recall the result of \cite{regev2009lattices}. This makes use of Gaussian-like lattice superposition states, which have the form 
\begin{equation}
\label{eqn-psi}
\ket{\psi} \propto \sum_{x\in L} \sum_{y\in\RR^n} f(y-x) \ket{y},
\end{equation}
where $L$ is a lattice in $\RR^n$, $f:\:\RR^n \rightarrow \RR$ is a radial function (such as a Gaussian), and for simplicity we are allowing the state to be unnormalized. (Later, we will show how to normalize $\ket{\psi}$, by restricting $x$ to a finite subset of $L$.) There are two questions: how to make use of $\ket{\psi}$, and how to prepare $\ket{\psi}$? We will address both of these.

Given polynomially many copies of $\ket{\psi}$, one can solve BDD with some decoding radius $r_\text{overlap}$. In fact, one can do more: one can measure $\ket{\psi}$ in the Fourier basis, to generate classical samples, which give a classical description of an oracle $O_{r_\text{overlap}}$ that solves BDD with decoding radius $r_\text{overlap}$. Here, $r_\text{overlap}$ is the largest distance such that, if $\ket{\phi}$ is a copy of $\ket{\psi}$ shifted by distance $r_\text{overlap}$, then the overlap $\innerprod{\phi}{\psi}$ is non-negligible. (Here we are assuming that $\lambda_1(L) = 1$, to keep the notation simple. This holds without any loss of generality.)

Unfortunately, preparing the state $\ket{\psi}$ seems to be hard. The obvious approach is to prepare the state 
\begin{equation}
\label{eqn-psi-index}
\ket{\psi_\text{index}} \propto \sum_{x\in L} \ket{x} \sum_{y\in\RR^n} f(y-x) \ket{y},
\end{equation}
and then ``erase'' or ``uncompute'' $x$, to get $\ket{0}\ket{\psi}$. Preparing $\ket{\psi_\text{index}}$ is easy, but erasing $x$ seems to be hard --- in particular, it seems to require solving BDD, with some decoding radius $r_\text{all}$. Here, $r_\text{all}$ is the smallest distance such that, for all but a negligible fraction of the terms $\ket{x}\ket{y}$ in eq.~(\ref{eqn-psi-index}), $\norm{y-x}_2 \leq r_\text{all}$.

If $r_\text{overlap} > r_\text{all}$, then the above techniques would lead to a quantum algorithm for solving BDD, by preparing states $\ket{\psi}$, and constructing oracles $O_r$ for solving BDD, with larger and larger decoding radii $r$ (note that BDD becomes harder when the decoding radius is larger). But unfortunately, this is not the case. For the most natural choice of the function $f$, namely a Gaussian, we have $r_\text{overlap} < r_\text{all}$, and more precisely, $r_\text{overlap} \sim r_\text{all}/\sqrt{n}$. (See Figure \ref{fig-radii}$(a)$.)

In \cite{regev2009lattices}, the LWE oracle is used to overcome this obstacle, by boosting the strength of the oracle $O_{r_\text{overlap}}$, i.e., by constructing a stronger oracle $O_{r_\text{lwe}}$ that solves BDD with a larger decoding radius $r_\text{lwe} > r_\text{overlap}$. By choosing the parameters of the LWE oracle appropriately, one can make $r_\text{lwe} > r_\text{all}$, and then repeat the procedure, to solve BDD with larger and larger decoding radii. (See Figure \ref{fig-radii}$(b)$.)

Here we study an alternate strategy, proposed in \cite{liu2010talk}, which attempts to use the quantum curvelet transform to overcome this obstacle, by weakening the oracle $O_{r_\text{all}}$ that is used to prepare the state $\ket{\psi}$, i.e., one attempts to prepare the state $\ket{\psi}$ using a weaker oracle $O_{r_\text{curv}}$ that solves BDD with a smaller decoding radius $r_\text{curv} < r_\text{all}$. (See Figure \ref{fig-radii}$(c)$.) This relies crucially on the behavior of the curvelet transform that was described above. 

This strategy is as follows: we apply the curvelet transform on the second register of the state $\ket{\psi_\text{index}}$; this returns a superposition of states $\ket{a,b,\theta}$ such that the line $\ell(b,\theta)$ passes near the point $x$. If the distance from $\ell(b,\theta)$ to $x$ is less than $r_\text{curv}$, then we can use the oracle $O_{r_\text{curv}}$ to find $x$, erase it, and prepare the state $\ket{\psi}$. If we can make $r_\text{curv} < r_\text{overlap}$, then this strategy will succeed in solving BDD, with larger and larger decoding radii.

In this paper, we prove that there is \textit{no} choice of the function $f$ in eq.~(\ref{eqn-psi}) (subject to a mild assumption about $f$), that will allow this strategy to succeed, i.e., by making $r_\text{curv} < r_\text{overlap}$. More precisely, for any radial function $f:\: \RR^n \rightarrow \RR$ that can be approximated (up to some negligible error) by a function whose support lies within a ball of radius $\lambda_1(L)/4$ around the origin, we prove the following bound:
\begin{equation}
\label{eqn-r-ratio}
\frac{r_\text{overlap}}{r_\text{curv}} \leq O(\sqrt{(\log n)/n}).
\end{equation}
This shows that $r_\text{overlap}$ is always too small, by a factor of $\sim 1/\sqrt{n}$. This rules out a natural class of quantum algorithms for solving lattice problems. 

This suggests that, for these kinds of algorithms to have any chance of success, they must use entangled measurements on multiple copies of the quantum state, perhaps along the lines of Kuperberg's sieve \cite{childs2010quantum}, or they must exploit some additional structure in the problem, perhaps by restricting to a special class of lattices, such as ideal lattices \cite{cramer2017short}. 

This result follows from lower-bounds on $r_\text{curv}$ (using our uncertainty principle for the curvelet transform), and upper-bounds on $r_\text{overlap}$ (using a nontrivial fact from harmonic analysis: for any radial function $f$ on $\RR^n$, in the limit as $n$ grows large, the autocorrelation function of $f$ can be approximated by a mixture of Gaussians).


\subsection{Outline of the Paper}
\label{sec-outline}

In the remainder of this paper, we present detailed proofs of the above results. First we present a new construction of the quantum curvelet transform, which simplifies many aspects of the previous work \cite{liu2009quantum}, in Section \ref{sec-curvelets}. We then prove our uncertainty relation (\ref{eqn-curvelet-uncertainty}) for the curvelet transform, in Section \ref{sec-uncertainty}. We prove the corresponding upper-bound (\ref{eqn-curvelet-upper-bound}) in Appendix \ref{sec-finding-center-upper-bounds}. 

We describe how the quantum curvelet transform might be used to prepare the lattice superposition state $\ket{\psi}$, in Section \ref{sec-obstacles}. Then we prove the infeasibility of this approach (in particular, eq.~(\ref{eqn-r-ratio})), in Sections \ref{sec-obstacles-2} and \ref{sec-obstacles-proof-details}. This provides the details omitted from the preceding Section \ref{subsec-obstacles-solving-lattice-problems}. Finally, we discuss some future research directions and related work in Section \ref{sec-outlook}.


\section{Quantum Curvelet Transforms}
\label{sec-curvelets}

In this section, we define a family of quantum curvelet transforms that will be the subject of this paper. First, we sketch how discrete curvelet transforms can be implemented on a quantum computer. Then we define continuous curvelet transforms, which can be viewed as the ideal mathematical objects that one seeks to approximate using discrete curvelet transforms. Continuous curvelet transforms are more convenient for theoretical analysis, and will be used to prove the main results of this paper.

As we will see below, our definition of a continuous curvelet transform is different from previous work \cite{liu2009quantum}. We will show that our definition preserves the desired properties of the curvelet transform (e.g., for finding the wavefront set), while avoiding some of the technical complications encountered in previous work.

\subsection{Discrete Curvelet Transforms on a Quantum Computer}

We define a family of discrete curvelet transforms that can be implemented on a quantum computer. The goal of this construction is to approximate the behavior of the continuous curvelet transform $\Gamma_f$, but acting on quantum superposition states in a finite-dimensional Hilbert space. This is related to classical algorithms for fast discrete curvelet transforms \cite{candes2006fast}, in the same way that the quantum Fourier transform is related to the classical fast Fourier transform.

The discrete curvelet transform is an isometry $T_\text{curvelet}$ that is implemented by a quantum circuit of the form:
\begin{equation*}
\Qcircuit @C=1em @R=1em {
&\ustick{x}\qw &\gate{QFT} &\ustick{k}\qw &\ctrl{1} &\ctrl{1} &\gate{QFT^{-1}} &\ustick{b}\qw &\qw \\
\lstick{\ket{0}} &\qw &\qw &\qw &\gate{W} &\ctrl{1} &\qw &\ustick{a}\qw &\qw \\
\lstick{\ket{0}} &\qw &\qw &\qw &\qw &\gate{V} &\qw &\ustick{\theta}\qw &\qw
}
\end{equation*}
This quantum circuit acts on three registers, which store positions $x\in\RR^n$ (or $b\in\RR^n$), scales $a\in(0,1)$, and directions $\theta\in S^{n-1}$, respectively. Each register has a finite-dimensional Hilbert space, so the above variables are restricted to discrete, finite domains. 

Initially, the first register contains the input to $T_\text{curvelet}$, and the second and third registers are prepared in fixed states $\ket{0}$ and $\ket{0}$. (To simplify the notation, we use $\ket{0}$ to denote both of these states, even though they belong to different Hilbert spaces.) Then the quantum circuit performs the following operations:
\begin{enumerate}
\item It applies a quantum Fourier transform (denoted $QFT$) on the first register, which produces a superposition $\ket{\hat{f}} := \sum_k \hat{f}(k) \ket{k}$, where $\hat{f}$ is the discrete Fourier transform of $f$. There are different ways to define this Fourier transform. For example, one can restrict the domain of $f$ to be a finite set of grid points inside a large cube in $\RR^n$, with periodic boundary conditions. Then one can view $f$ as a function on the additive group $(\mathbb{Z}_M)^n$, and apply the Fourier transform over this group.
\item It applies a controlled-$W$ operation that maps $\ket{k}\ket{0}$ to $\ket{k}\sum_a W_{k,a} \ket{a}$. For each point $k$ in the frequency domain, this prepares a superposition of different scales $a$. The coefficients $W_{k,a}$ are chosen to approximate the continuous curvelet transform, which we will define in the next section. In particular, the coefficients are chosen so that, for each scale $a$, the mapping $k\mapsto W_{k,a}$ is an approximation of the corresponding radial window function.
\item It applies a doubly-controlled-$V$ operation that maps $\ket{k}\ket{a}\ket{0}$ to $\ket{k}\ket{a}\sum_\theta V_{k,a,\theta} \ket{\theta}$. For each point $k$ in the frequency domain, and each scale $a$, this prepares a superposition of different directions $\theta$. The coefficients $V_{k,a,\theta}$ are chosen to approximate the continuous curvelet transform, which we will define in the next section. In particular, the coefficients are chosen so that, for each scale $a$ and direction $\theta$, the mapping $k\mapsto V_{k,a,\theta}$ is an approximation of the corresponding angular window function.
\item It applies the inverse quantum Fourier transform $QFT^{-1}$ on the first register. This produces a superposition over different positions $b$.
\end{enumerate}


\subsection{Continuous Curvelet Transforms}

We now define a family of continuous curvelet transforms $\Gamma_f$. These are intended to be the ideal mathematical functions that are approximated by the discrete curvelet transforms defined in the previous section. Let $f:\: \RR^n \rightarrow \RR$, and define 
\begin{equation}
\Gamma_f(a,b,\theta) = \int_{\RR^n} \hat{f}(k) \chi_{a,\theta}(k) e^{2\pi ik\cdot b} dk,
\end{equation}
which takes arguments $0<a\leq 1$, $b\in\RR^n$ and $\theta\in S^{n-1}$, where $S^{n-1}$ is the unit sphere in $\RR^n$. Here $\hat{f}$ is the (continuous) Fourier transform of $f$, and $\chi_{a,\theta}:\: \RR^n \rightarrow \RR$ is a ``window function'' that is defined over the frequency domain (described below). This is the continuous analogue of the discrete curvelet transform $U_{DCT}$ defined in the previous section.

$\Gamma_f$ can also be written in an equivalent form:
\begin{equation}
\label{eqn-curvelet-def-2}
\Gamma_f(a,b,\theta) = \int_{\RR^n} \gamma_{a,b,\theta}(x)^* f(x) dx,
\end{equation}
where $\gamma_{a,b,\theta}:\: \RR^n \rightarrow \CC$ are curvelet basis functions, and $^*$ denotes the complex conjugate. Here, $\gamma_{a,b,\theta}$ is the translation of $\gamma_{a,0,\theta}$ to position $b$, and $\gamma_{a,0,\theta}$ is the inverse Fourier transform of the window function $\chi_{a,\theta}$, that is, 
\begin{equation}
\label{eqn-curvelet-basis}
\gamma_{a,b,\theta}(x) = \gamma_{a,0,\theta}(x-b), \quad
\hat{\gamma}_{a,0,\theta}(k) = \chi_{a,\theta}(k).
\end{equation}

To construct the function $\chi_{a,\theta}$, we write $k\in\RR^n$ using spherical coordinates centered around the vector $\theta$. That is, we use the mapping
\begin{multline}
\label{eqn-spherical-coords}
k \mapsto (r,\phi_1,\ldots,\phi_{n-2},\phi_{n-1}) \\
\in [0,\infty) \times [0,\pi]^{n-2} \times [0,2\pi),
\end{multline}
where $r = \norm{k}_2$, $\phi_1$ is chosen so that $k\cdot\theta = r\cos\phi_1$, and the remaining angles $\phi_2,\ldots,\phi_{n-1}$ are spherical coordinates in the subspace of $\RR^n$ orthogonal to $\theta$. 

We let $\chi_{a,\theta}$ be a product of a ``radial'' window function that depends on $r$, and is large when $r\sim 1/a$; and an ``angular'' window function that depends on $\phi_1$, and is large when $\phi_1 \lesssim \sqrt{a}$. This is motivated by the following geometric picture: the support of $\chi_{a,\theta}$ is a subset of the frequency domain, with thickness $\sim 1/a$ in the direction $\theta$, and width $\sim 1/\sqrt{a}$ in the directions orthogonal to $\theta$. If one takes the inverse Fourier transform of $\chi_{a,\theta}$, one gets a curvelet basis function $\gamma_{a,0,\theta}$, which has thickness $\sim a$ in the direction $\theta$, and width $\sim \sqrt{a}$ in the directions orthogonal to $\theta$.

More precisely, we define $\chi_{a,\theta}$ to be:
\begin{equation}
\label{eqn-chi}
\chi_{a,\theta}(k) = W(\lambda ar) V(\tfrac{1}{\sqrt{a}} \sin^+\phi_1) C_{a,n}.
\end{equation}
Here, the functions $W$ and $V$ play the same role as the controlled-$W$ and controlled-$V$ operations in the discrete curvelet transform $U_{DCT}$. The radial window function $W$ can be any function $W:\: [0,\infty) \rightarrow [0,\infty)$ that is supported on the interval $[\tfrac{1}{e}, 1]$, and satisfies the admissibility condition 
\begin{equation}
\int_0^\infty W(r)^2 \frac{dr}{r} = 1.
\end{equation}
For simplicity, one can set $\lambda = 1$; or one can choose different values $\lambda \neq 1$, in order to expand or contract the support of $\chi_{a,\theta}$, and thereby expand or contract the support of $\gamma_{a,b,\theta}$. This adjusts the definition of the scale $a$ in terms of the frequency variable $k$ (or equivalently, the position variable $x$).

The angular window function $V$ can be any function $V:\: [0,\infty) \rightarrow [0,\infty)$ that is supported on the interval $[0, \tfrac{1}{\sqrt{2}}]$ and is monotone decreasing. The function $\sin^+:\: [0,\pi] \rightarrow [0,1]$ is defined to be 
\begin{equation}
\sin^+\phi_1 = \begin{cases}
\sin\phi_1 &\text{if } 0 \leq \phi_1 \leq \pi/2, \\
1 &\text{if } \pi/2 < \phi_1 \leq \pi.
\end{cases}
\end{equation}
The normalization factor $C_{a,n}$ is chosen so that 
\begin{equation}
\int_{S^{n-1}} V(\tfrac{1}{\sqrt{a}} \sin^+\phi_1)^2 C_{a,n}^2 d\sigma(\phi) = 1,
\end{equation}
where $d\sigma$ denotes integration over $S^{n-1}$ using spherical coordinates. 

The normalization factor $C_{a,n}$ can be bounded by:
\begin{equation}
\label{eqn-normalization-bounds}
M_{n-2} a^{(n-1)/2} S'_0 \leq C_{a,n}^{-2} \leq \sqrt{2} M_{n-2} a^{(n-1)/2} S'_0,
\end{equation}
where we define
\begin{equation}
\label{eqn-M-n-2}
M_{n-2} = \int_0^{1/\sqrt{2}} V(y)^2 y^{n-2} dy,
\end{equation}
\begin{equation}
S'_0 = \int_{S^{n-2}} d\sigma(\phi_2,\ldots,\phi_{n-1}).
\end{equation}
(This follows from an elementary calculation: write 
\begin{equation}
C_{a,n}^{-2} = \int_0^{\sin^{-1}(\sqrt{a/2})} V(\tfrac{1}{\sqrt{a}} \sin\phi_1)^2 \sin^{n-2}\phi_1 d\phi_1 S'_0,
\end{equation}
use the bounds $\cos\phi_1 \leq 1 \leq \sqrt{2}\cos\phi_1$ inside the integral, and change variables $y = \tfrac{1}{\sqrt{a}} \sin\phi_1$.)

Using these definitions, it is straightforward to verify that $\chi_{a,\theta}$ satisfies the following identity: for all $k\in\RR^n$ such that $\norm{k}_2 \geq 1/\lambda$, 
\begin{equation}
\label{eqn-curvelet-window-identity}
\int_0^1 \int_{S^{n-1}} \chi_{a,\theta}(k)^2 d\sigma(\theta) \frac{da}{a} = 1. 
\end{equation}

This implies that the curvelet transform $\Gamma_f$ has the following important properties. Assume that $\hat{f}(k) = 0$ whenever $\norm{k}_2 < 1/\lambda$, i.e., $\hat{f}$ vanishes below some low-frequency cutoff $1/\lambda$. Then $f$ can be reconstructed from $\Gamma_f$, using the formula:
\begin{equation}
\label{eqn-curvelet-inversion}
f(x) = \int_{S^{n-1}} \int_{\RR^n} \int_0^1 \Gamma_f(a,b,\theta) \gamma_{a,b,\theta}(x) \frac{da}{a} db d\sigma(\theta).
\end{equation}

In addition, $\Gamma_f$ satisfies an analogue of Plancherel's theorem:
\begin{equation}
\label{eqn-curvelet-plancherel}
\int_{\RR^n} \abs{f(x)}^2 dx = \int_{S^{n-1}} \int_{\RR^n} \int_0^1 \abs{\Gamma_f(a,b,\theta)}^2 \frac{da}{a} db d\sigma(\theta).
\end{equation}
This allows us to view $\Gamma_f$ as a properly-normalized quantum wavefunction $\ket{\Gamma_f}$, assuming that $f$ is properly normalized. The integral $\int_0^1 da/a$ may look unusual, but it can be written in a more pedestrian form $\int_{-\infty}^0 ds$, by changing variables $s = \log a$. 

Finally, for any fixed scale $a\leq 1$, $\Gamma_f$ satisfies the following variant of Plancherel's theorem, where we restrict $\hat{f}$ to the subset of the frequency domain that corresponds to $a$:
\begin{multline}
\label{eqn-curvelet-plancherel-a}
\int_{\RR^n} \abs{\hat{f}(k)}^2 W(\lambda a\norm{k}_2)^2 dk \\
= \int_{S^{n-1}} \int_{\RR^n} \abs{\Gamma_f(a,b,\theta)}^2 db d\sigma(\theta).
\end{multline}

The proofs of eqs.~(\ref{eqn-curvelet-window-identity})-(\ref{eqn-curvelet-plancherel-a}) are the same as in previous work \cite{candes2005continuous, liu2009quantum}, except for two minor differences. First, we use integrals with respect to the measure $(da/a) db d\sigma(\theta)$, rather than $(da/a^{n+1}) db d\sigma(\theta)$. This choice of the measure makes no difference for computing the expectation values of measurements of the wavefunction $\ket{\Gamma_f}$. Hence, in this paper, we choose $(da/a) db d\sigma(\theta)$ for simplicity. However, we note that identical results can be obtained using the measure $(da/a^{n+1}) db d\sigma(\theta)$. This choice of the measure is useful for constructing the discrete curvelet transform, where one samples the function $\Gamma_f$ at the Nyquist rate, which depends on $a$.

Second, we define the angular window function $V$ to depend on $\tfrac{1}{\sqrt{a}} \sin^+\phi_1$, rather than $\phi_1/\sqrt{a}$. This leads to simpler calculations, since it is more compatible with the measure $\sin^{n-2} \phi_1 d\phi_1$, when integrating using spherical coordinates in $\RR^n$. In particular, this removes the need for the complicated ``adjustment factor'' $\Lambda_a(\phi_1)$ that was used in \cite{liu2009quantum}. 

Due to the above change, our version of the curvelet transform behaves slightly differently from the version of the curvelet transform studied previously in \cite{liu2009quantum}. However, our upper-bound in eq.~(\ref{eqn-curvelet-upper-bound}) suggests that our version of the curvelet transform can be used for all of the applications described in \cite{liu2009quantum}. 


\subsection{Transforms for Computing the Scale and Direction}
\label{subsec-scale-direction}

For certain calculations, it is convenient to view the curvelet transform as a composition of two separate operations that compute the ``scale'' and ``direction.''

In the discrete case, the curvelet transform is an isometry $T_\text{curvelet}$ that maps $\ket{f} = \sum_x f(x) \ket{x}$ to a superposition over states $\ket{a}\ket{b}\ket{\theta}$. $T_\text{curvelet}$ can be written as a product of two isometries, $T_\text{scale}$ and $T_\text{direction}$, which first compute the scales $a$, and then the directions $\theta$. That is, 
\begin{equation}
T_\text{curvelet} = T_\text{direction} T_\text{scale},
\end{equation}
where $T_\text{scale}$ maps $\ket{\psi}$ to a superposition over states $\ket{a}\ket{b}$, and $T_\text{direction}$ maps this to a superposition over states $\ket{a}\ket{b}\ket{\theta}$. $T_\text{scale}$ is defined by the quantum circuit:
\begin{equation*}
\Qcircuit @C=1em @R=1em {
&\ustick{x}\qw &\gate{QFT} &\ustick{k}\qw &\ctrl{1} &\gate{QFT^{-1}} &\ustick{x}\qw &\qw \\
\lstick{\ket{0}} &\qw &\qw &\qw &\gate{W} &\qw &\ustick{a}\qw &\qw
}
\end{equation*}
and $T_\text{direction}$ is defined by the quantum circuit:
\begin{equation*}
\Qcircuit @C=1em @R=1em {
&\ustick{x}\qw &\gate{QFT} &\ustick{k}\qw &\ctrl{1} &\gate{QFT^{-1}} &\ustick{b}\qw &\qw \\
&\ustick{a}\qw &\qw &\qw &\ctrl{1} &\qw &\ustick{a}\qw &\qw \\
\lstick{\ket{0}} &\qw &\qw &\qw &\gate{V} &\qw &\ustick{\theta}\qw &\qw
}
\end{equation*}

The continuous case can be handled in a similar way. First, define $\Xi_f$ to be the transform that computes the scale $a$,
\begin{equation}
\label{eqn-xi-f}
\Xi_f(a,x) = \int_{\RR^n} \hat{f}(k) W(\lambda a\norm{k}_2) e^{2\pi ik\cdot x} dk.
\end{equation}
Note that for each scale $a$, $\Xi_f(a,\cdot)$ can be interpreted as a quantum wavefunction, whose $L_2$ norm satisfies the following identities, due to eqs.~(\ref{eqn-curvelet-plancherel-a}) and (\ref{eqn-curvelet-plancherel}):
\begin{multline}
\label{eqn-xi-f-plancherel-a}
\int_{\RR^n} \abs{\Xi_f(a,x)}^2 dx
= \int_{S^{n-1}} \int_{\RR^n} \abs{\Gamma_f(a,b,\theta)}^2 db d\sigma(\theta),
\end{multline}
\begin{equation}
\label{eqn-xi-f-plancherel}
\int_0^1 \int_{\RR^n} \abs{\Xi_f(a,x)}^2 dx \frac{da}{a} = \int_{\RR^n} \abs{f(x)}^2 dx.
\end{equation}

Next, for each scale $a$, define $\Theta^{(a)}_f$ to be the transform that computes the direction $\theta$,
\begin{equation}
\label{eqn-curvelet-single-scale}
\Theta^{(a)}_f(b,\theta) = \int_{\RR^n} \hat{f}(k) \chi^{(a)}_\theta(k) e^{2\pi ik\cdot b} dk,
\end{equation}
\begin{equation}
\label{eqn-curvelet-window-single-scale}
\chi^{(a)}_\theta(k) = V(\tfrac{1}{\sqrt{a}} \sin^+\phi_1) C_{a,n}.
\end{equation}
Then the curvelet transform $\Gamma_f$ can be written in terms of $\Xi_f$ and $\Theta^{(a)}_f$, as follows:
\begin{equation}
\Gamma_f(a,b,\theta) = \Theta^{(a)}_{\Xi_f(a,\cdot)}(b,\theta).
\end{equation}


\section{An Uncertainty Principle for Curvelets}
\label{sec-uncertainty}

We now prove the uncertainty principle in eq.~(\ref{eqn-curvelet-uncertainty}). Without loss of generality, let $c = \vec{0} \in \RR^n$. We claim that there exist constants $C>0$ and $C'>0$ such that, for any radial function $f:\: \RR^n \rightarrow \RR$ centered at $\vec{0}$, and for any $0<a\leq 1$, 
\begin{equation}
\label{eqn-uncert-1}
\begin{split}
\int_{S^{n-1}} &\int_{\RR^n} b^T(I-\theta\theta^T)b\, \abs{\Gamma_f(a,b,\theta)}^2 db d\sigma(\theta) \\
&\geq Ca \int_{\RR^n} \norm{x}_2^2 \abs{\Xi_f(a,x)}^2 dx \\
&+ C'\lambda^2 n^2 a^2 \int_{S^{n-1}} \int_{\RR^n} \abs{\Gamma_f(a,b,\theta)}^2 db d\sigma(\theta).
\end{split}
\end{equation}
Here $\lambda$ is a parameter of the curvelet transform, which defines the scale $a$ with respect to the frequency domain (see eq.~(\ref{eqn-chi})). We will prove (\ref{eqn-uncert-1}) in several steps.


\subsection{Simplifying the Integrals}
\label{sec-simplifying-integrals}

Let $T$ denote the left-hand side of eq.~(\ref{eqn-uncert-1}), and let $0<a\leq 1$ be fixed. We can rewrite $T$ in terms of the direction and scale transforms (see Section \ref{subsec-scale-direction}) as follows:
\begin{equation}
T = \int_{S^{n-1}} \int_{\RR^n} b^T(I-\theta\theta^T)b \abs{\Theta^{(a)}_h(b,\theta)}^2 db d\sigma(\theta),
\end{equation}
where $h:\: \RR^n \rightarrow \RR$ is the function 
\begin{equation}\label{eqn-h-def}
h(x) = \Xi_f(a,x). 
\end{equation}
Notice that, since $f$ is a radial function, $h$ is real-valued, and $\hat{h}$ is a radial function:
\begin{equation}
\hat{h}(k) = \hat{f}(k) W(\lambda a\norm{k}_2).
\end{equation}
For convenience, we can write $\hat{h}$ in the form $\hat{h}(k) = H_0(\norm{k}_2)$, where $H_0:\: [0,\infty) \rightarrow \RR$; and similarly we can write $\hat{f}(k) = F_0(\norm{k}_2)$. Then $H_0$ and $F_0$ are related by:
\begin{equation}
\label{eqn-H0}
H_0(r) = F_0(r) W(\lambda ar).
\end{equation}

We can simplify $T$ by using the rotational symmetries of the problem, as described in \cite{liu2009quantum}. First, we can evaluate the integral over $\theta$, to get 
\begin{equation}
T = S_0 \int_{\RR^n} b^T(I-uu^T)b \abs{\Theta^{(a)}_h(b,u)}^2 db,
\end{equation}
where $S_0$ is the surface area of the unit sphere $S^{n-1}$ in $\RR^n$, and $u = (1,0,\ldots,0)$ is the first standard basis vector in $\RR^n$. Note that $\Theta^{(a)}_h$ has an additional symmetry 
\begin{equation}
\Theta^{(a)}_h(R(b),u) = \Theta^{(a)}_h(b,u), 
\end{equation}
where $R$ can be any rotation on $\RR^n$ such that $R(u) = u$. This implies
\begin{equation}
T = S_0 (n-1) \int_{\RR^n} b_2^2 \abs{\Theta^{(a)}_h(b,u)}^2 db.
\end{equation}
Finally, we can use Plancherel's theorem to rewrite the integral over the frequency domain:
\begin{equation}
T = \frac{S_0 (n-1)}{(2\pi)^2} \int_{\RR^n} \bigl\lvert \tfrac{\partial}{\partial k_2} \widehat{\Theta}^{(a)}_h(k,u) \bigr\rvert^2 dk,
\end{equation}
where 
\begin{equation}
\widehat{\Theta}^{(a)}_h(k,u) = \hat{h}(k) \chi^{(a)}_u(k),
\end{equation}
using eq.~(\ref{eqn-curvelet-single-scale}).

We now introduce spherical coordinates $k \mapsto (r,\phi_1,\ldots,\phi_{n-1})$, where $k\cdot u = r\cos\phi_1$, as in eq.~(\ref{eqn-spherical-coords}). We then have 
\begin{equation}
\widehat{\Theta}^{(a)}_h(k,u) = H_0(r) X_0(\phi_1),
\end{equation}
where $H_0$ is defined in eq.~(\ref{eqn-H0}), and we define the function $X_0:\: [0,\pi] \rightarrow [0,\infty)$ to be 
\begin{equation}
X_0(\phi_1) = V(\tfrac{1}{\sqrt{a}} \sin^+\phi_1) C_{a,n},
\end{equation}
as in eq.~(\ref{eqn-curvelet-window-single-scale}). Differentiating by $k_2$, we get 
\begin{multline}
\tfrac{\partial}{\partial k_2} \widehat{\Theta}^{(a)}_h(k,u) \\
= H'_0(r) X_0(\phi_1) \tfrac{\partial r}{\partial k_2} + H_0(r) X'_0(\phi_1) \tfrac{\partial\phi_1}{\partial k_2},
\end{multline}
where 
\begin{equation}
\tfrac{\partial r}{\partial k_2} = \sin\phi_1 \cos\phi_2, \quad
\tfrac{\partial\phi_1}{\partial k_2} = \tfrac{1}{r} \cos\phi_1 \cos\phi_2.
\end{equation}

We can now rewrite $T$ as a sum of three simpler integrals:
\begin{equation}
\label{eqn-T}
T = \frac{S_0 (n-1)}{(2\pi)^2} (I_A + 2I_B + I_C),
\end{equation}
where
\begin{equation}
I_A = \int_{S^{n-1}} \int_0^\infty \Bigl( H'_0(r) X_0(\phi_1) \tfrac{\partial r}{\partial k_2} \Bigr)^2 r^{n-1} dr d\sigma(\phi),
\end{equation}
\begin{multline}
I_B = \int_{S^{n-1}} \int_0^\infty H'_0(r) X_0(\phi_1) \tfrac{\partial r}{\partial k_2} \cdot \\
H_0(r) X'_0(\phi_1) \tfrac{\partial\phi_1}{\partial k_2} r^{n-1} dr d\sigma(\phi),
\end{multline}
\begin{equation}
I_C = \int_{S^{n-1}} \int_0^\infty \Bigl( H_0(r) X'_0(\phi_1) \tfrac{\partial\phi_1}{\partial k_2} \Bigr)^2 r^{n-1} dr d\sigma(\phi).
\end{equation}
We will use the trivial lower-bound $I_C \geq 0$. We will then prove nontrivial lower-bounds for $I_B$ and $I_A$.


\subsection{Lower-Bounding $I_B$}
\label{sec-lower-bounding-IB}

We can write $I_B$ as a product of three factors $I_{Br}$, $I_{B1}$ and $I_2$, which are integrals over $r$, $\phi_1$ and $(\phi_2,\ldots,\phi_{n-1})$, as follows:
\begin{equation}
\label{eqn-IB}
I_B = I_{Br} I_{B1} I_2,
\end{equation}
\begin{equation}
I_{Br} = \int_0^\infty H'_0(r) H_0(r) r^{n-2} dr,
\end{equation}
\begin{equation}
I_{B1} = \int_0^\pi X_0(\phi_1) X'_0(\phi_1) \cos\phi_1 \sin^{n-1}\phi_1 d\phi_1,
\end{equation}
\begin{equation}
\label{eqn-I2}
I_2 = \int_{S^{n-2}} \cos^2\phi_2 d\sigma(\phi_2,\ldots,\phi_{n-1}).
\end{equation}

To evaluate $I_2$, we use the identities
\begin{equation}
\begin{split}
\int_0^\pi &\cos^2\phi_2 \sin^{n-3}\phi_2 d\phi_2 
= \frac{1}{n-2} \int_0^\pi \sin^{n-1}\phi_2 d\phi_2 \\
&= \int_0^\pi \sin^{n-3}\phi_2 d\phi_2 - \int_0^\pi \sin^{n-1}\phi_2 d\phi_2,
\end{split}
\end{equation}
where the first line follows from integration by parts, and the second line follows from the identity $\cos^2\phi_2 = 1 - \sin^2\phi_2$. After some algebra, we get an exact expression for $I_2$:
\begin{equation}
\label{eqn-I2-calc}
I_2 = \frac{S'_0}{n-1}, \quad S'_0 = \int_{S^{n-2}} d\sigma(\phi_2,\ldots,\phi_{n-1}).
\end{equation}

To bound $I_{B1}$, recall that the angular window function $V$ is supported on $[0, \tfrac{1}{\sqrt{2}}]$, so we have 
\begin{equation}
\begin{split}
I_{B1} &= \int_0^{\sin^{-1}(\sqrt{a/2})} V(\tfrac{1}{\sqrt{a}} \sin\phi_1) V'(\tfrac{1}{\sqrt{a}} \sin\phi_1) \cdot \\
&\qquad\qquad\qquad\quad C_{a,n}^2 \tfrac{1}{\sqrt{a}} \cos^2\phi_1 \sin^{n-1}\phi_1 d\phi_1.
\end{split}
\end{equation}
Inside the integral, we have $V'(\tfrac{1}{\sqrt{a}} \sin\phi_1) \leq 0$, and $\cos\phi_1 \geq \tfrac{1}{\sqrt{2}}$, hence 
\begin{equation}
\begin{split}
I_{B1} &\leq \tfrac{1}{\sqrt{2}} \int_0^{\sin^{-1}(\sqrt{a/2})} V(\tfrac{1}{\sqrt{a}} \sin\phi_1) V'(\tfrac{1}{\sqrt{a}} \sin\phi_1) \cdot \\
&\qquad\qquad\qquad\qquad\quad C_{a,n}^2 \tfrac{1}{\sqrt{a}} \cos\phi_1 \sin^{n-1}\phi_1 d\phi_1.
\end{split}
\end{equation}
After changing variables (letting $y = \tfrac{1}{\sqrt{a}} \sin\phi_1$), integrating by parts, and using eq.~(\ref{eqn-normalization-bounds}) to bound $C_{a,n}^2$, we get
\begin{equation}
\label{eqn-IB1-calc}
\begin{split}
I_{B1}
&\leq -C_{a,n}^2 \tfrac{1}{2\sqrt{2}} (n-1) a^{(n-1)/2} \int_0^{1/\sqrt{2}} V(y)^2 y^{n-2} dy \\
&\leq -\frac{n-1}{4S'_0}.
\end{split}
\end{equation}

To bound $I_{Br}$, recall that $H_0$ is supported on $[\tfrac{1}{\lambda ae}, \tfrac{1}{\lambda a}]$. Then we can integrate by parts to get 
\begin{equation}
\label{eqn-IBr-calc}
\begin{split}
I_{Br} &= -\tfrac{1}{2} (n-2) \int_0^\infty H_0(r)^2 r^{n-3} dr \\
&\leq -\tfrac{1}{2} (n-2) \lambda^2 a^2 \int_0^\infty F_0(r)^2 W(\lambda ar)^2 r^{n-1} dr.
\end{split}
\end{equation}

Substituting into eq.~(\ref{eqn-IB}), we get
\begin{equation}
I_B \geq \tfrac{1}{8} (n-2) \lambda^2 a^2 \int_0^\infty F_0(r)^2 W(\lambda ar)^2 r^{n-1} dr.
\end{equation}
Hence we have:
\begin{multline}
\tfrac{S_0 (n-1)}{2\pi^2} I_B \geq \tfrac{1}{16\pi^2} S_0 (n-1) (n-2) \lambda^2 a^2 \cdot \\
\int_0^\infty F_0(r)^2 W(\lambda ar)^2 r^{n-1} dr.
\end{multline}
Then using eq.~(\ref{eqn-curvelet-plancherel-a}), we get
\begin{multline}
\tfrac{S_0 (n-1)}{2\pi^2} I_B \geq \tfrac{1}{16\pi^2} (n-1) (n-2) \lambda^2 a^2 \cdot \\
\int_{S^{n-1}} \int_{\RR^n} \abs{\Gamma_f(a,b,\theta)}^2 db d\sigma(\theta).
\end{multline}
We substitute this into eq.~(\ref{eqn-T}), to lower-bound $T$.


\subsection{Lower-Bounding $I_A$}
\label{sec-lower-bounding-IA}

As we did with $I_B$, we can write $I_A$ as a product of three factors:
\begin{equation}
\label{eqn-IA}
I_A = I_{Ar} I_{A1} I_2,
\end{equation}
\begin{equation}
I_{Ar} = \int_0^\infty H'_0(r)^2 r^{n-1} dr,
\end{equation}
\begin{equation}
I_{A1} = \int_0^\pi X_0(\phi_1)^2 \sin^n\phi_1 d\phi_1,
\end{equation}
where $I_2 = \frac{S'_0}{n-1}$, as in eqs.~(\ref{eqn-I2}) and (\ref{eqn-I2-calc}).

To bound $I_{A1}$, recall that the angular window function $V$ is supported on $[0, \tfrac{1}{\sqrt{2}}]$, so we have 
\begin{equation}
\begin{split}
I_{A1} &= \int_0^{\sin^{-1}(\sqrt{a/2})} V(\tfrac{1}{\sqrt{a}} \sin\phi_1)^2 
C_{a,n}^2 \sin^n\phi_1 d\phi_1.
\end{split}
\end{equation}
Inside the integral, we have $\cos\phi_1 \leq 1$, and $\sin\phi_1 \geq 0$, hence 
\begin{equation}
\begin{split}
I_{A1} &\geq \int_0^{\sin^{-1}(\sqrt{a/2})} V(\tfrac{1}{\sqrt{a}} \sin\phi_1)^2
C_{a,n}^2 \cos\phi_1 \sin^n\phi_1 d\phi_1.
\end{split}
\end{equation}
After changing variables (letting $y = \tfrac{1}{\sqrt{a}} \sin\phi_1$), and using eq.~(\ref{eqn-normalization-bounds}) to bound $C_{a,n}^2$, we get
\begin{equation}
\label{eqn-I-A1-calc}
\begin{split}
I_{A1}
&\geq C_{a,n}^2 a^{(n+1)/2} \int_0^{1/\sqrt{2}} V(y)^2 y^n dy \\
&\geq \frac{M_n}{M_{n-2}} \frac{a}{\sqrt{2}S'_0},
\end{split}
\end{equation}
where $M_n$ and $M_{n-2}$ are defined as in eq.~(\ref{eqn-M-n-2}).

The Cauchy-Schwartz inequality implies that 
\begin{equation}
M_{n-2} \leq \sqrt{M_n} \sqrt{M_{n-4}},
\end{equation}
which implies 
\begin{equation}
\frac{M_n}{M_{n-2}} \geq \frac{M_{n-2}}{M_{n-4}} \geq \cdots 
\geq \min\biggl\lbrace \frac{M_2}{M_0}, \frac{M_3}{M_1} \biggr\rbrace.
\end{equation}
Plugging into (\ref{eqn-I-A1-calc}), we get
\begin{equation}
I_{A1} \geq \min\bigl\lbrace \tfrac{M_2}{M_0}, \tfrac{M_3}{M_1} \bigr\rbrace \frac{a}{\sqrt{2}S'_0}.
\end{equation}

In order to compute $I_{Ar}$, we will evaluate the following integral involving the function $h(x) = \Xi_f(a,x)$ (see eq.~(\ref{eqn-h-def})), using similar techniques as above (i.e., the radial symmetry of $h$, Plancherel's theorem, and spherical coordinates):
\begin{equation}
\begin{split}
\int_{\RR^n} &\norm{x}_2^2 \abs{h(x)}^2 dx
= \frac{n}{(2\pi)^2} \int_{\RR^n} \abs{\tfrac{\partial}{\partial k_1} \hat{h}(k)}^2 dk \\
&= \frac{n}{(2\pi)^2} \int_{S^{n-1}} \int_0^\infty (H'_0(r) \tfrac{\partial r}{\partial k_1})^2 r^{n-1} dr d\sigma(\phi),
\end{split}
\end{equation}
where
\begin{equation}
\tfrac{\partial r}{\partial k_1} = \cos\phi_1.
\end{equation}
We can evaluate the integral over $\phi$, 
\begin{equation}
\int_{S^{n-1}} \cos^2\phi_1 d\sigma(\phi) = \frac{S_0}{n},
\end{equation}
in the same way that we computed $I_2$ in eq.~(\ref{eqn-I2-calc}). Then we get
\begin{equation}
\int_{\RR^n} \norm{x}_2^2 \abs{h(x)}^2 dx
= \frac{S_0}{(2\pi)^2} \int_0^\infty H'_0(r)^2 r^{n-1} dr.
\end{equation}
This gives us the following expression for $I_{Ar}$:
\begin{equation}
\label{eqn-IAr-calc}
I_{Ar} = \frac{(2\pi)^2}{S_0} \int_{\RR^n} \norm{x}_2^2 \abs{h(x)}^2 dx.
\end{equation}

Substituting into eq.~(\ref{eqn-IA}), we get
\begin{equation}
I_A \geq \tfrac{(2\pi)^2}{S_0} 
\min\bigl\lbrace \tfrac{M_2}{M_0}, \tfrac{M_3}{M_1} \bigr\rbrace \tfrac{a}{\sqrt{2}(n-1)} 
\int_{\RR^n} \norm{x}_2^2 \abs{h(x)}^2 dx.
\end{equation}
Hence we have:
\begin{equation}
\tfrac{S_0 (n-1)}{(2\pi)^2} I_A \geq
\min\bigl\lbrace \tfrac{M_2}{M_0}, \tfrac{M_3}{M_1} \bigr\rbrace \tfrac{a}{\sqrt{2}} 
\int_{\RR^n} \norm{x}_2^2 \abs{h(x)}^2 dx.
\end{equation}
We substitute this into eq.~(\ref{eqn-T}), to lower-bound $T$.


\section{Preparing Lattice Superposition States}
\label{sec-obstacles}

In the following two sections, we give a detailed description of the result sketched in Section \ref{subsec-obstacles-solving-lattice-problems}, showing that a certain class of quantum algorithms for solving lattice problems cannot succeed. In this section, we describe a possible approach to solving lattice problems on a quantum computer, by preparing lattice superposition states using the quantum curvelet transform. In Sections \ref{sec-obstacles-2} and \ref{sec-obstacles-proof-details}, we prove that this approach cannot succeed, for a large class of lattice superposition states.


\subsection{Lattice Superposition States}
\label{subsec-defining-psi}

We will begin by defining Gaussian-like lattice superposition states, and explaining their relevance to solving radius-$r$ BDD (with $r$ scaling inverse-polynomially in $n$), and $\gamma$-approximate SVP (with $\gamma$ scaling polynomially with $n$).

Let $L = \mathcal{L}(b_1,\ldots,b_n)$ be a lattice in $\RR^n$. Without loss of generality, we will assume that the basis $b_1,\ldots,b_n$ has been reduced using the LLL algorithm, which runs in polynomial time \cite{nguyen2010lll}. We will work with a finite subset of lattice points $Q\subset L$, where 
\begin{equation}
\label{eqn-Q}
Q = \set{ \sum_{j=1}^n v_j b_j \;|\; v_1,\ldots,v_n \in \set{-M,-M+1,\ldots,M} },
\end{equation}
and $M$ is an integer. 

We will choose the precise value of $M$ later, in Section \ref{sec-Q}. In general, we will allow $M$ to be exponentially large as a function of the number of bits needed to describe the basis vectors $b_1,\ldots,b_n$. In particular, $M$ may be exponentially large in the dimension $n$. Even in this regime, there can still be efficient algorithms for preparing complex quantum superpositions over $Q$, using techniques whose running time is polynomial in $\log M$, such as quantum sampling, the quantum Fourier transform, and the quantum curvelet transform \cite{grover2002creating, nielsen2000quantum, liu2009quantum}. 

Our goal is to construct a quantum state $\ket{\psi}$ that is a superposition over points $y\in\RR^n$, and has larger amplitude when $y$ is close to a lattice point $x\in Q$. We define $\ket{\psi}$ as follows:
\begin{equation}
\label{eqn-lattice-Gaussians}
\ket{\psi} = \frac{1}{Z} \sum_{x\in Q} \sum_{y\in\RR^n} f(y-x) \ket{y},
\end{equation}
where $Z$ is a normalization factor, and $f:\: \RR^n \rightarrow \RR$ is a radial function centered at 0, that is, $f$ can be written in the form $f(x) = f_0(\norm{x}_2)$, for some function $f_0:\: [0,\infty) \rightarrow \RR$. Typically, $f_0$ is chosen to be a monotonically decreasing, Gaussian-like function. This gives a larger amplitude to those points $y$ that are near lattice points $x\in Q$. 

Note that we have abused the notation by writing $\ket{\psi}$ as a summation over $y$, rather than an integral. This imitates the notation used to describe quantum algorithms. This is justified because $\ket{\psi}$ can be approximated by a state in a finite-dimensional Hilbert space, by restricting $y$ to a discrete grid, and imposing periodic boundary conditions. Then the resulting state can be represented on a quantum computer. 

The state $\ket{\psi}$ can be used to solve BDD, with decoding radius $r\sim 1/\sqrt{n}$, in the following way. Suppose that the vector $t$ is within distance $r\lambda_1(L)$ of a lattice point $x\in L$. Define $S_t$ to be the unitary operator that shifts a point in $\RR^n$ by $t$, that is, 
\begin{equation}
S_t:\: \ket{y} \mapsto \ket{y+t}, \; \forall y\in\RR^n.
\end{equation}
Then the overlap $\bra{\psi}S_t\ket{\psi}$ between $\ket{\psi}$ and $S_t\ket{\psi}$ is non-negligible (i.e., it is at least $1/\text{poly}(n)$), and it increases monotonically as $t$ approaches the lattice point $x$. By measuring the state $\ket{\psi}$, and estimating this overlap, one can estimate the value of the function $t\mapsto \bra{\psi}S_t\ket{\psi}$. Then, starting from $t$, one can perform gradient ascent to find the lattice point $x$. 

In order for the above reasoning to hold, we must choose the set $Q$ in eq.~(\ref{eqn-lattice-Gaussians}) appropriately, so that $t$ is deep in the interior of the convex hull of $Q$, and hence the overlap $\bra{\psi}S_t\ket{\psi}$ is not affected by the fact that the superposition in $\ket{\psi}$ only involves a finite subset of the lattice. This can be accomplished using standard techniques (e.g., by reducing $t$ using Babai's nearest-plane algorithm \cite{babai1986lovasz}), as described in Section \ref{sec-Q}. 

In addition to solving radius-$r$ BDD, the state $\ket{\psi}$ can also be used to solve $\gamma$-approximate SVP. This follows from the fact that the quantum Fourier transform of $\ket{\psi}$ is a superposition over short vectors in the dual lattice $L^*$. This technique is described in \cite{regev2009lattices}. 

Furthermore, using this technique, the computation of the overlap $\bra{\psi}S_t\ket{\psi}$ can be ``de-quantized,'' in the following sense \cite{aharonov2005lattice, regev2009lattices}. One can measure the quantum Fourier transform of $\ket{\psi}$ to obtain classical data, i.e., short vectors in $L^*$ sampled at random from a certain distribution. These random samples can be used to estimate the overlap $\bra{\psi}S_t\ket{\psi}$ at any point $t$, up to some additive error $\pm\epsilon$. 

This ``de-quantized'' procedure for estimating $\bra{\psi}S_t\ket{\psi}$ has two important properties, which are useful for upper-bounding the complexity of lattice problems. First, the same random samples in $L^*$ can be re-used many times, to evaluate $\bra{\psi}S_t\ket{\psi}$ at many different points $t$. This is unlike the quantum state $\ket{\psi}$, which can only be used once, since it is destroyed when it is measured. Second, if one wants to evaluate $\bra{\psi}S_t\ket{\psi}$ at polynomially many points $t$, with inverse-polynomial additive error, then it is sufficient to use polynomially many samples from $L^*$, i.e., the number of samples scales polynomially with $n$, $1/\epsilon$, and the number of bits needed to describe the basis vectors $b_1,\ldots,b_n$. 

However, the above reasoning does not imply a polynomial-time quantum algorithm for radius-$r$ BDD, or for $\gamma$-approximate SVP, because there is an important caveat: there is no known way to prepare the state $\ket{\psi}$ in polynomial time. We will discuss this issue in the following section.


\subsection{Preparing the State $\ket{\psi}$}
\label{subsec-preparing-psi}

To see the difficulty in preparing the lattice superposition state $\ket{\psi}$ in eq.~(\ref{eqn-lattice-Gaussians}), consider one straightforward approach to performing this task: 
\begin{enumerate}
\item Use standard techniques \cite{grover2002creating} to prepare the state 
\begin{equation}
\label{eqn-state-prep-easy}
\frac{1}{Z'} \sum_{x\in Q} \ket{x} \sum_{y\in\RR^n} f(y) \ket{y},
\end{equation}
with a running time that scales polynomially in $\log M$.
\item Apply a shift operation $S_x$, controlled by $x$, that maps $\ket{y}$ to $\ket{y+x}$, in order to get the state 
\begin{equation}
\label{eqn-index-erasure}
\frac{1}{Z'} \sum_{x\in Q} \ket{x} \sum_{y\in\RR^n} f(y-x) \ket{y} =: \ket{\psi_\text{index}}.
\end{equation}
We call this state $\ket{\psi_\text{index}}$, because the contents of the second register resembles the desired superposition $\ket{\psi}$, but the second register is entangled with the first register, i.e., different terms in the superposition are labeled or ``indexed'' by different states $\ket{x}$ in the first register.
\item Apply an operation, controlled by $y$, to ``uncompute'' $\ket{x}$ from the first register. This disentangles the two registers, and produces the desired state: 
\begin{equation}
\frac{1}{Z} \sum_{x\in Q} \ket{0} \sum_{y\in\RR^n} f(y-x) \ket{y} = \ket{0}\ket{\psi}.
\end{equation}
\end{enumerate}

This last step is a special case of a more general problem called ``index erasure'' \cite{ambainis2011symmetry}. It is not clear how to perform this step, because one needs to erase $x$ while preserving the superposition over $y$. There is also some evidence that this task might be hard, although this evidence is not conclusive. 

First, if this task is performed using only classical computation (i.e., using only classical gates that permute the standard basis states), this seems to require computing $x$ from $y$, which is as hard as solving BDD in the first place. But if this task is performed using quantum gates, which can produce superpositions of the standard basis states, it is unclear whether this task is still hard. 

Second, this task can also be viewed as a concrete instantiation of an abstract oracle problem, whose quantum query complexity can be lower bounded \cite{ambainis2011symmetry}. But these lower bounds do not apply to concrete instantiations of the oracle problem.

The quantum reduction from $\gamma$-approximate SVP to LWE \cite{regev2009lattices} gives a new perspective on this question. Essentially, it shows that one \textit{can} perform the ``index erasure'' step after eq.~(\ref{eqn-index-erasure}), under certain circumstances --- in particular, by using an oracle that solves the LWE problem. This implies that the query complexity of performing index erasure, by itself, is \textit{not} a sufficient obstacle to prevent one from preparing the state $\ket{\psi}$. This motivates the question: could there be \textit{other} ways to perform index erasure, without using an LWE oracle? 

In fact, there is another plausible way of doing this index erasure operation, using the quantum curvelet transform, without an LWE oracle \cite{liu2010talk}. We will describe this next.


\subsection{Index Erasure Using the Quantum Curvelet Transform}
\label{subsec-index-erasure-curvelet-trans}

We now describe a possible approach to preparing the state $\ket{\psi}$, where index erasure is done via the quantum curvelet transform \cite{liu2010talk}. To see this, first recall how index erasure is done using the LWE oracle, following \cite{regev2009lattices} (though here we are using somewhat different notation):
\begin{enumerate}
\item Given the state $\ket{\psi_\text{index}}$ in eq.~(\ref{eqn-index-erasure}), let $r_\text{all}$ be the smallest real number such that, for almost all terms $\ket{x}\ket{y}$ in the superposition $\ket{\psi_\text{index}}$ (excluding a subset of terms $\ket{x}\ket{y}$ whose total probability is negligible), we have $\norm{y-x}_2 \leq r_\text{all}\lambda_1(L)$. 
\item Suppose we have an oracle $O_{r_\text{all}}$ that solves BDD with decoding radius $r_\text{all}$. Then we can run $O_{r_\text{all}}$ on $y$ to find $x$, erase it from the first register, and obtain the state $\ket{0}\ket{\psi}$. 
\item The state $\ket{\psi}$ can be used to solve BDD with some decoding radius $r_\text{overlap} < r_\text{all}$ (specifically, $r_\text{overlap} \sim r_\text{all}/\sqrt{n}$). Then one can use classical random sampling techniques (i.e., the ``de-quantized'' procedure for estimating the overlap $\bra{\psi} S_t \ket{\psi}$ in Section \ref{subsec-defining-psi}) to construct an oracle $O_{r_\text{overlap}}$ that solves BDD with decoding radius $r_\text{overlap}$ (without requiring more copies of $\ket{\psi}$). 
\item One can use the LWE oracle to construct a \textit{stronger} oracle $O_{r_\text{lwe}}$ that solves BDD with decoding radius $r_\text{lwe} > r_\text{overlap}$ (specifically, $r_\text{lwe} \sim \alpha pr_\text{overlap}$, where $\alpha$ and $p$ are parameters of the LWE oracle, defined in \cite{regev2009lattices}).
\item If $r_\text{lwe}/r_\text{all} \geq 1 + \tfrac{1}{\text{poly}(n)}$, then one can apply this procedure repeatedly, to construct BDD oracles with larger and larger decoding radii, up to $\sim 1/\sqrt{n}$.
\end{enumerate}
Essentially, the above procedure works because in step 4, the LWE oracle is able to increase the decoding radius by a factor of $\alpha p$, which overcomes the loss in step 3, where the decoding radius decreases by a factor of $\sqrt{n}$.

Now consider a different approach to performing index erasure, using the quantum curvelet transform, without an LWE oracle \cite{liu2010talk}: 
\begin{enumerate}
\item As before, given the state $\ket{\psi_\text{index}}$ in eq.~(\ref{eqn-index-erasure}), let $r_\text{all}$ be the smallest real number such that, for almost all terms $\ket{x}\ket{y}$ in the superposition $\ket{\psi_\text{index}}$, we have $\norm{y-x}_2 \leq r_\text{all}\lambda_1(L)$. 

\item Apply the quantum curvelet transform to the register containing $\ket{y}$. This produces a superposition of states $\ket{a,b,\theta}$. For almost all terms in the superposition, the line $\ell(b,\theta)$ (see eq.~(\ref{eqn-line})) passes near the center $x$ of the radial function $y\mapsto f(y-x)$. 

\item Suppose one has an oracle $O_{r_\text{curv}}$ that solves BDD with decoding radius $r_\text{curv}$, where $\frac{1}{\text{poly}(n)} r_\text{all} \leq r_\text{curv} < r_\text{all}$. Furthermore, suppose the line $\ell(b,\theta)$ passes within distance $\frac{1}{\sqrt{2}} r_\text{curv}\lambda_1(L)$ of $x$. One can use the procedure described below, in order to find $x$. One can then erase $x$ from the first register, and obtain the state $\ket{0}\ket{\psi}$.
\begin{enumerate}
\item Let $\epsilon = \frac{1}{\sqrt{2}} r_\text{curv}\lambda_1(L)$, and let $S_\epsilon$ be an $\epsilon$-net covering an appropriate subset $S$ of $\ell(b,\theta)$ (discussed below), such that $\abs{S_\epsilon} \leq \text{poly}(n)$, and $S_\epsilon$ contains at least one point $\tilde{x}$ that lies within distance $\sqrt{2}\epsilon = r_\text{curv}\lambda_1(L)$ of $x$. Then run the oracle $O_{r_\text{curv}}$ on every point $\tilde{x}$ in $S_\epsilon$. On one of these points, the oracle will return $x$. 
\end{enumerate}

\item Suppose that $\ket{\psi}$ can be used to solve BDD with decoding radius $r_\text{overlap} < r_\text{all}$. Then one can use the same techniques as above, to construct an oracle $O_{r_\text{overlap}}$ that solves BDD with decoding radius $r_\text{overlap}$ (without requiring more copies of $\ket{\psi}$). 

\item If $r_\text{overlap}/r_\text{curv} \geq 1 + \tfrac{1}{\text{poly}(n)}$, then one can proceed iteratively as before, to construct BDD oracles with larger and larger decoding radii, up to $\sim 1/\sqrt{n}$.
\end{enumerate}

\begin{figure}
\includegraphics{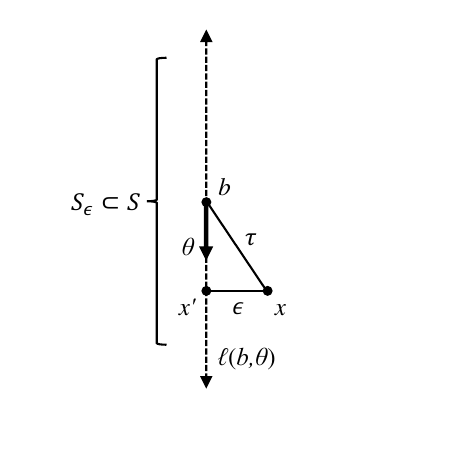}
\caption{Finding a point $x$, given a point $b$ and a direction $\theta$. We are promised that the line $\ell(b,\theta)$ passes within distance $\epsilon$ of the point $x$, and that $\norm{b-x}_2 \leq \tau$. $S$ is a line segment, lying along $\ell(b,\theta)$, of length $2\tau$, centered at $b$. $S_\epsilon$ is an $\epsilon$-net for $S$. Then $S_\epsilon$ contains a point within distance $\sqrt{2}\epsilon$ of $x$.}
\label{fig-line}
\end{figure}

In the above procedure, the sets $S_\epsilon \subset S \subset \ell(b,\theta)$ can be constructed as follows (see Fig.~\ref{fig-line}):
\begin{enumerate}
\item Suppose we have an upper-bound $\tau$ on $\norm{b-x}_2$, that is, 
\begin{equation}
\norm{b-x}_2 \leq \tau, 
\end{equation}
such that 
\begin{equation}
\tau \leq \text{poly}(n) r_\text{all} \lambda_1(L).
\end{equation}
Typically, such an upper-bound $\tau$ can be proved using the fact that the curvelet transform $\Gamma_f(a,b,\theta)$ (fixing $a$ and $\theta$) is the convolution of $f$ with the curvelet basis element $\gamma_{a,\theta}$ (see eq.~(\ref{eqn-curvelet-def-2})); the fact that $f$ is supported within radius $r_\text{all}$; the fact that $\gamma_{a,\theta}$ is a rapidly decaying function, since its Fourier transform is smooth (see eq.~(\ref{eqn-curvelet-basis})); and standard results about concentration of measure when the dimension $n$ is large \cite{ball1997elementary}. 

\item Then let 
\begin{multline}
S = \set{\tilde{x} \in \ell(b,\theta) \;|\; \norm{\tilde{x}-b}_2 \leq \tau},
\end{multline}
and let $S_\epsilon$ be an $\epsilon$-net for $S$. It is straightforward to check that $S_\epsilon$ has the desired properties. First, 
\begin{equation}
\abs{S_\epsilon} \leq 2\tau/\epsilon \leq \text{poly}(n) r_\text{all} / r_\text{curv} \leq \text{poly}(n). 
\end{equation}
Second, $S$ contains the point 
\begin{equation}
x' = \arg\min_{\tilde{x}\in \ell(b,\theta)} \norm{\tilde{x}-x}_2, 
\end{equation}
which satisfies $\norm{x'-x}_2 \leq \epsilon$. Hence $S_\epsilon$ contains at least one point $\tilde{x}$ such that $\norm{\tilde{x}-x}_2 \leq \sqrt{2}\epsilon$.
\end{enumerate}

The above procedure will succeed if one can choose the function $f$ in eq.~(\ref{eqn-lattice-Gaussians}) so that the quantum curvelet transform gives accurate results that make $r_\text{curv}$ small in step 3, while keeping $r_\text{overlap}$ large in step 4, so that the algorithm can make progress towards solving BDD with larger and larger decoding radii. (See the earlier discussion in Fig.~\ref{fig-radii}.)

Some upper bounds on $r_\text{curv}$ are known, for some specific choices of the function $f$ \cite{liu2009quantum}. But these upper bounds are too large, by roughly a factor of $\sqrt{n}$ relative to $r_\text{overlap}$, to make the above procedure succeed. However, this leaves open the possibility that a better choice of the function $f$, and stronger techniques for upper bounding the error in step 1, might be able to close this $\sqrt{n}$ gap, and succeed in solving BDD with decoding radius $\sim 1/\text{poly}(n)$. 

In this paper, we rule out this possibility, by showing nearly-tight \textit{lower bounds} on $r_\text{curv}$, using our uncertainty principle for the quantum curvelet transform. These lower bounds imply that the above approach cannot succeed in solving $\gamma$-approximate SVP, regardless of the choice of the function $f$ in eq.~(\ref{eqn-lattice-Gaussians}). We describe this in the next section.


\section{An Obstacle to Solving BDD}
\label{sec-obstacles-2}

In order for the above algorithm to succeed in solving BDD, the radial function $f$ in eq.~(\ref{eqn-lattice-Gaussians}) should be chosen to make $r_\text{overlap}$ large and $r_\text{curv}$ small. This means that $f$ has to satisfy two requirements: 
\begin{enumerate}
\item[$(i)$] $f$ should be smooth enough, so that the overlap $\bra{\psi}S_t\ket{\psi}$ is large when the vector $t$ is at distance $r_\text{overlap}$ from the lattice;
\item[$(ii)$] $f$ should have wave-like oscillations, so that one can use the curvelet transform $\Gamma_f$ to find a line $\ell(b,\theta)$ that passes within distance $r_\text{curv}$ of the nearest lattice point. 
\end{enumerate}

There is a potential conflict between these two requirements, because they imply that, at high frequencies (i.e., when $\norm{k}_2$ is large), the Fourier transform $\hat{f}(k)$ must decay rapidly (in order to satisfy $(i)$), but not too rapidly (in order to satisfy $(ii)$). This conflict is apparent in some of the examples studied in \cite{liu2009quantum}. For instance, when $f$ is the indicator function of a ball, then it satisfies $(i)$ but not $(ii)$; and when $f$ is the indicator function of a thin spherical shell, then it satisfies $(ii)$ but not $(i)$.

In this paper we show that this conflict is unavoidable. More precisely, for any choice of the radial function $f$ (subject to a mild assumption), we show that there is a tradeoff between satisfying the requirements $(i)$ and $(ii)$, i.e., it is impossible to satisfy both requirements simultaneously. Formally, we prove the bound stated in eq.~(\ref{eqn-r-ratio}):
\begin{equation}
\frac{r_\text{overlap}}{r_\text{curv}} \leq O(\sqrt{(\log n)/n}).
\end{equation}
This constitutes an obstacle to solving BDD, using the approach described in section \ref{subsec-index-erasure-curvelet-trans}.

Here we sketch the overall argument, while deferring some of the detailed calculations to Section \ref{sec-obstacles-proof-details}. First, we assume that the support of $f$ is contained inside a ball around the origin, whose radius is strictly less than $\tfrac{1}{4} \lambda_1(L)$. For typical choices of $f$, this assumption is satisfied up to a small error, which becomes negligible as the dimension $n$ grows large. Without loss of generality, we let $f$ be normalized so that its $L_2$ norm is equal to 1. 

Next we consider requirement $(i)$, and prove an upper-bound on $r_\text{overlap}$. Then we consider requirement $(ii)$, and prove a lower-bound on $r_\text{curv}$.


\subsection{Upper-Bounding $r_\text{overlap}$}

Regarding requirement $(i)$, we show that for any target vector $t$ whose distance from the lattice is strictly less than $\tfrac{1}{2} \lambda_1(L)$, the overlap $\bra{\psi}S_t\ket{\psi}$ can be approximated by:
\begin{equation}
\label{eqn-overlap-0}
\bra{\psi}S_t\ket{\psi} \approx \bra{f}S_s\ket{f},
\end{equation}
where $s = t-w$ is the difference between the target vector $t$ and the nearest lattice point $w$, and $\ket{f}$ is the quantum state whose wavefunction is given by $f$. For the proof of eq.~(\ref{eqn-overlap-0}), see Section \ref{sec-Q}.

Next, we use the curvelet transform to decompose $\ket{f}$ into contributions $\ket{f^{(a)}}$ at different scales $a$. The curvelet transform can be written as an isometry $T_\text{curvelet}$ that maps $\ket{f}$ to a superposition over states $\ket{a}\ket{b}\ket{\theta}$, where $a$, $b$ and $\theta$ represent scales, locations and directions, respectively. Recall from Section \ref{subsec-scale-direction} that $T_\text{curvelet}$ can be written as a product of two isometries, $T_\text{scale}$ and $T_\text{direction}$, which first compute the scales $a$, and then the directions $\theta$. That is, 
\begin{equation}
\label{eqn-curvelet-two-steps}
T_\text{curvelet} = T_\text{direction} T_\text{scale},
\end{equation}
where $T_\text{scale}$ maps $\ket{\psi}$ to a superposition over states $\ket{a}\ket{b}$, and $T_\text{direction}$ maps this to a superposition over states $\ket{a}\ket{b}\ket{\theta}$. 

We interpret $T_\text{scale} \ket{f}$ as a decomposition of $\ket{f}$ into contributions $\ket{f^{(a)}}$ at different scales $a$, with complex amplitudes $q_a$, that is:
\begin{equation}
\label{eqn-f-decomposition}
T_\text{scale} \ket{f} = \sum_a q_a \ket{a}\ket{f^{(a)}},
\end{equation}
where $\sum_a \abs{q_a}^2 = 1$ and $\innerprod{f^{(a)}}{f^{(a)}} = 1$. This decomposition has the property that each state $\ket{f^{(a)}}$ is band-limited over the frequency domain, i.e., its Fourier transform $QFT \ket{f^{(a)}}$ is a superposition of states $\ket{k}$ where $k$ lies in the annulus 
\begin{equation}
\label{eqn-annulus}
\set{k\in\RR^n \;|\; \tfrac{1}{e\lambda a} \leq \norm{k}_2 \leq \tfrac{1}{\lambda a}}
\end{equation}
(see Section \ref{subsec-uncertainty}).

In addition, $T_\text{direction}$ has a simple structure: it is an isometry that is controlled by the $a$ register and acts on the $b$ register, i.e., $T_\text{direction}$ is block-diagonal, 
\begin{equation}
\label{eqn-curvelet-blocks}
T_\text{direction} = \bigoplus_a \ket{a}\bra{a} \tensor T_\text{direction}^{(a)},
\end{equation}
where $T_\text{direction}^{(a)}$ is the block that corresponds to the state $\ket{a}$. Hence we can write the curvelet transform of $\ket{f}$ in terms of the states $\ket{f^{(a)}}$, as follows:
\begin{equation}
\label{eqn-curvelet-psi-decomposition}
T_\text{curvelet} \ket{f} = \sum_a q_a \ket{a} T_\text{direction}^{(a)} \ket{f^{(a)}}.
\end{equation}
This says that when we apply the curvelet transform to the state $\ket{f}$, $T_\text{direction}$ acts independently on each of the states $\ket{f^{(a)}}$. 

Turning once more to requirement $(i)$, we show that the overlap $\bra{f}S_s\ket{f}$ is a probabilistic mixture of the overlaps involving the states $\ket{f^{(a)}}$, that is, 
\begin{equation}
\label{eqn-overlap-lb-0}
\bra{f}S_s\ket{f} = \sum_a \abs{q_a}^2 \bra{f^{(a)}}S_s\ket{f^{(a)}}.
\end{equation}
(This follows because the operator $T_\text{scale}$ is block-diagonal, and the operator $S_s$ is diagonal, with respect to the Fourier basis.) Furthermore, we show that 
\begin{multline}
\label{eqn-overlap-lb}
\bra{f}S_s\ket{f} \\
\approx \sum_a \abs{q_a}^2 \EE_{k\sim QFT\ket{f^{(a)}}}\biggl[ \exp\biggl( -\frac{2\pi^2 \norm{s}_2^2\norm{k}_2^2}{n-2} \biggr) \biggr],
\end{multline}
where $\EE_k[\cdot]$ denotes averaging over random measurement outcomes $k$ obtained by measuring the quantum Fourier transform of $\ket{f^{(a)}}$, and the approximation is accurate up to an additive error of size $\exp(-\Omega(n))$. For the proof of eq.~(\ref{eqn-overlap-lb}), see Section \ref{subsec-autocorrelation}.

This can be viewed as a generalization of a fact used in previous works involving Gaussian-like lattice superposition states: when the states $\ket{\psi}$ are constructed using a radial function $f$ that is a Gaussian, the overlap $\bra{\psi}S_t\ket{\psi}$ behaves like a Gaussian function of $s$ \cite{regev2009lattices}. Eqs.~(\ref{eqn-overlap-0}) and (\ref{eqn-overlap-lb}) show that, when the states $\ket{\psi}$ are constructed using a radial function $f$ that is chosen arbitrarily (i.e., not necessarily Gaussian), the overlap $\bra{\psi}S_t\ket{\psi}$ can still be approximated by a \textit{mixture} of Gaussian functions of $s$. 

Combining eqs.~(\ref{eqn-overlap-0}) and (\ref{eqn-overlap-lb}) with the bound $\norm{k}_2 \geq 1/(e\lambda a)$ from eq.~(\ref{eqn-annulus}), we get that the overlap $\bra{\psi}S_t\ket{\psi}$ is upper-bounded by
\begin{multline}
\label{eqn-overlap-lb-1}
\bra{\psi}S_t\ket{\psi} \leq 
\sum_a \abs{q_a}^2 \exp\biggl( -\frac{2\pi^2 \norm{s}_2^2}{e^2\lambda^2 a^2(n-2)} \biggr) \\
+ \exp(-\Omega(n)).
\end{multline}
In the above sum, the contribution from scale $a$ becomes negligible when $\norm{s}_2 \geq \omega(\sqrt{n\log n} \lambda a)$. This implies that the state $\ket{\psi}$ can be used to solve BDD with decoding radius at most
\begin{equation}
\label{eqn-overlap-lb-2}
r_\text{overlap} \leq O(\sqrt{n\log n} \lambda a_\text{max}),
\end{equation}
where $a_\text{max}$ is the largest value of $a$ that appears (with a non-negligible amplitude $q_a$) in the decomposition of $\ket{f}$ in eq.~(\ref{eqn-f-decomposition}).


\subsection{Lower-Bounding $r_\text{curv}$}

Next, with regard to requirement $(ii)$, consider what happens when we apply the quantum curvelet transform to the second register of the state $\ket{\psi_\text{index}}$. From the output of the curvelet transform, we construct the line $\ell(b,\theta)$. We are interested in estimating the expectation value of $d(\ell(b,\theta),x)^2$, the square of the distance from the line $\ell(b,\theta)$ to the point $x$ encoded in the first register of the quantum state. This is given by the expectation value of the observable
\begin{equation}
H = \sum_{x\in Q} \sum_{a,b,\theta} d(\ell(b,\theta),x)^2 \ket{x}\bra{x} \tensor \ket{a,b,\theta} \bra{a,b,\theta}.
\end{equation}

A straightforward calculation shows that 
\begin{equation}
\label{eqn-center-finding-lb-0}
\begin{split}
\bra{\psi_\text{index}} &(I\tensor T_\text{curvelet}) H (I\tensor T_\text{curvelet}) \ket{\psi_\text{index}} \\
&= \EE_{(a,b,\theta) \sim T_\text{curvelet} \ket{f}} (b^T(I-\theta\theta^T)b),
\end{split}
\end{equation}
where we used the fact that  
\begin{equation}
\ket{\psi_\text{index}} = \frac{1}{Z'} \sum_{x\in Q} (S_x \tensor S_x) \ket{0}\ket{f};
\end{equation}
we used the fact that the operator $T_\text{curvelet}$ is block-diagonal, and the operator $S_x$ is diagonal, with respect to the Fourier basis; 
and $\EE_{a,b,\theta}[\cdot]$ denotes an average over random measurement outcomes obtained by measuring the state $T_\text{curvelet} \ket{f}$. 

Decomposing $\ket{f}$ into states $\ket{f^{(a)}}$ (see eq.~(\ref{eqn-f-decomposition})), and using (the second term of) our uncertainty relation for the curvelet transform (\ref{eqn-curvelet-uncertainty}), we can lower-bound this quantity as follows:
\begin{equation}
\label{eqn-center-finding-lb}
\begin{split}
&\EE_{(a,b,\theta)\sim T_\text{curvelet} \ket{f}} (b^T(I-\theta\theta^T)b) \\
&\qquad= \sum_a \abs{q_a}^2 \EE_{(b,\theta)\sim T_\text{direction}^{(a)} \ket{f^{(a)}}} (b^T(I-\theta\theta^T)b) \\
&\qquad\geq \Omega\Bigl( \sum_a \abs{q_a}^2 \lambda^2 n^2 a^2 \Bigr).
\end{split}
\end{equation}

Now combine eqs.~(\ref{eqn-center-finding-lb-0}) and (\ref{eqn-center-finding-lb}), and use this to analyze the algorithm in Section \ref{subsec-preparing-psi}. In the algorithm, we have access to an oracle $O_{r_\text{curv}}$ that solves BDD with decoding radius $r_\text{curv}$. For every $(a,b,\theta)$ in the superposition state $(I\tensor T_\text{curvelet}) \ket{\psi_\text{index}}$, we construct a set of points $S_\epsilon$ that lie on the line $\ell(b,\theta)$, such that $S_\epsilon$ contains at least one point $\tilde{x}$ with $\norm{\tilde{x}-x}_2 \leq \sqrt{2} d(\ell(b,\theta),x)$. 

Then we run the oracle $O_{r_\text{curv}}$ on every point $\tilde{x}$ in $S_\epsilon$. On at least one of these points $\tilde{x}$, the oracle will succeed in finding $x$. Then we erase $x$ from the first register, and apply $T_\text{curvelet}^{-1}$ on the second register, in order to produce the state $\ket{0}\ket{\psi}$. For this strategy to succeed, we need an oracle $O_{r_\text{curv}}$ that solves BDD with decoding radius 
\begin{equation}
\label{eqn-center-finding-lb-2}
r_\text{curv} \geq \Omega(\lambda na_\text{max}),
\end{equation}
where $a_\text{max}$ was defined previously in eq.~(\ref{eqn-overlap-lb-2}).

Together, eqs.~(\ref{eqn-overlap-lb-2}) and (\ref{eqn-center-finding-lb-2}) imply that $r_\text{overlap}/r_\text{curv} \leq O(\sqrt{(\log n)/n})$, for any choice of the radial function $f$. This proves eq.~(\ref{eqn-r-ratio}). This is below the threshold $r_\text{overlap}/r_\text{curv} \geq 1 + \tfrac{1}{\text{poly}(n)}$ that is needed for the algorithm described in the previous section to succeed. Hence, this is an obstacle to solving BDD using the quantum curvelet transform.


\section{Obstacle to Solving BDD: Technical Details}
\label{sec-obstacles-proof-details}


In this section, we prove two technical results that were used in Section \ref{sec-obstacles-2}, to analyze the overlap $\bra{\psi}S_t\ket{\psi}$: eqs.~(\ref{eqn-overlap-0}) and (\ref{eqn-overlap-lb}). 



\subsection{A Finite Subset of Lattice Points, and Proof of Eq.~(\ref{eqn-overlap-0})}
\label{sec-Q}

First, we address one technical issue: we need to set the parameter $M$ that defines the finite subset of lattice points $Q$ in eq.~(\ref{eqn-Q}), which is used to construct the superposition state $\ket{\psi}$ in eq.~(\ref{eqn-lattice-Gaussians}). In particular, we need to choose $M$ sufficiently large, so that when using the state $\ket{\psi}$ to solve BDD (as discussed in Section \ref{subsec-defining-psi}), the target vector $t$ lies deep inside the convex hull of $Q$, and hence the overlap $\bra{\psi}S_t\ket{\psi}$ is not significantly affected by the finite extent of the set $Q$. This will enable us to prove eq.~(\ref{eqn-overlap-0}).

Suppose that our goal is to solve BDD on a lattice $L = \mathcal{L}(b_1,\ldots,b_n)$ with target vector $t$, i.e., we are promised that there exists a lattice point $w\in L$ such that $\norm{w-t}_2 < \tfrac{1}{2} \lambda_1(L)$.

Without loss of generality, we can assume that $\norm{t}_2 \leq 2^{Cn}\lambda_1(L)$, where $C$ is some universal constant. (This holds because we can run Babai's nearest plane algorithm \cite{babai1986lovasz}, to find a point $\tilde{w}\in L$ that is within distance $2^{O(n)}\lambda_1(L)$ of $t$. Then we can reduce our problem to one with a new target vector $t-\tilde{w}$, which is close to the lattice point $w-\tilde{w}$.) 

Now consider the set $Q$ in eq.~(\ref{eqn-Q}), which is defined using the parameter $M$. We will set $M$ as follows. Suppose that the basis vectors $b_1,\ldots,b_n$ can be written using fixed-point numbers with $\ell$ bits of precision. Without loss of generality, we can rescale the lattice so that every basis vector $b_j$ has integer coordinates in the range $\set{-2^\ell,-2^\ell+1,\ldots,2^\ell}$. Then set 
\begin{equation}
M = 2^\kappa nM', \quad M' = \lceil (2^{Cn} + \tfrac{1}{2}) (2n) 2^{\ell n+n\log_2 n} \rceil, 
\end{equation}
where $\kappa \geq 1$ is an integer which we will adjust later. 

Consider the state $\ket{\psi}$ in eq.~(\ref{eqn-lattice-Gaussians}), which is defined using a radial function $f$. Without loss of generality, let $f$ be normalized so that its $L_2$ norm is equal to 1. 

The following claim holds for any choice of $f$, such that the support of $f$ is contained inside a ball around the origin, whose radius is strictly less than $\tfrac{1}{4} \lambda_1(L)$. We claim that the overlap $\bra{\psi}S_t\ket{\psi}$ satisfies the following bounds:
\begin{equation}
\label{eqn-overlap}
(1-2^{-\kappa-1}) \zeta(s) \leq \bra{\psi}S_t\ket{\psi} \leq \zeta(s),
\end{equation}
where $s = t-w$ is the difference between the target vector $t$ and the nearest lattice point $w$; and $\zeta:\: \RR^n \rightarrow \RR$ is the autocorrelation function of $f$, 
\begin{equation}
\label{eqn-zeta}
\zeta(z) = \bra{f}S_z\ket{f} = \int_{\RR^n} f(y)f(y-z)dy.
\end{equation}

Eq.~(\ref{eqn-overlap}) shows that the overlap $\bra{\psi}S_t\ket{\psi}$ is approximately equal to $\zeta(s)$, up to a multiplicative factor of $(1-2^{-\kappa-1})$, which becomes exponentially close to 1 as one increases the parameter $\kappa$. Note that we can choose $\kappa$ to be polynomially large, and the quantum algorithms for state preparation used in eq.~(\ref{eqn-state-prep-easy}) will still be efficient, since their running time scales polynomially with $\log M$, which scales polynomially with $n$, $\ell$ and $\kappa$. Hence this multiplicative factor of $(1-2^{-\kappa-1})$ can essentially be ignored, for the purposes of designing a polynomial-time quantum algorithm. In particular, this implies eq.~(\ref{eqn-overlap-0}).

We now prove eq.~(\ref{eqn-overlap}). We can write the state $\ket{\psi}$ in eq.~(\ref{eqn-lattice-Gaussians}) as follows:
\begin{equation}
\ket{\psi} = \frac{1}{(2M+1)^{n/2}} \sum_{x\in Q} \sum_{y\in G+x} f(y-x) \ket{y}.
\end{equation}
This can be written more properly as a wavefunction $\psi(y)$, since $y$ takes on continuous values in $\RR^n$. However, here and in eq.~(\ref{eqn-lattice-Gaussians}), we write $\ket{\psi}$ as a sum of basis states $\ket{y}$, in order to imitate the notation used when $y$ is restricted to a discrete set of grid points, so that the state $\ket{\psi}$ can be represented on a quantum computer. 

A straightforward calculation shows that
\begin{equation}
S_t\ket{\psi} = \frac{1}{(2M+1)^{n/2}} \sum_{x\in Q+w} \sum_{y\in G+x} f(y-x) \ket{y+s},
\end{equation}
and since $\norm{s}_2 \leq r\lambda_1(L)$, and $2r' + r < 1$, we have
\begin{equation}
\label{eqn-overlap-2}
\bra{\psi}S_t\ket{\psi} = \frac{\abs{Q\cap (Q+w)}}{(2M+1)^n} \zeta(s).
\end{equation}
It is obvious that $\abs{Q\cap (Q+w)} \leq \abs{Q} = (2M+1)^n$, and this proves the second part of eq.~(\ref{eqn-overlap}).

To prove the first part of eq.~(\ref{eqn-overlap}), we need to lower-bound $\abs{Q\cap (Q+w)}$. To do this, let $B \in \RR^{n\times n}$ be the matrix whose columns are the basis vectors $b_1,\ldots,b_n$, and write $w = Bv$ for some vector $v \in \RR^n$. Without loss of generality, we can assume that $B$ is full-rank, hence $v$ is uniquely defined. Then we have 
\begin{equation}
\abs{Q\cap (Q+w)} \geq (2M+1-\norm{v}_\infty)^n.
\end{equation}

Next we use the upper-bound 
\begin{equation}
\norm{v}_\infty \leq \norm{v}_2 \leq \norm{B^{-1}} \norm{w}_2.
\end{equation}
Using Minkowski's second theorem \cite{micciancio2002complexity}, and Stirling's approximation, we have 
\begin{equation}
\begin{split}
\norm{w}_2
&< (2^{Cn}+\tfrac{1}{2}) \lambda_1(L) \leq (2^{Cn}+\tfrac{1}{2}) \abs{\det(B)}^{1/n} \\
&\leq (2^{Cn}+\tfrac{1}{2}) (n!)^{1/n} 2^\ell < (2^{Cn}+\tfrac{1}{2}) (2n) 2^\ell.
\end{split}
\end{equation}

Let $\sigma_1(B) \geq \cdots \geq \sigma_n(B) > 0$ denote the singular values of the matrix $B$, and let $\norm{B}$ and $\norm{B}_F$ denote the operator norm and the Frobenius norm of $B$. Using the fact that $\abs{\det(B)} \geq 1$ (since $B$ is nonsingular with integer entries), we have 
\begin{equation}
\begin{split}
\norm{B^{-1}}
&= \frac{1}{\sigma_n(B)} = \frac{\prod_{j=1}^{n-1} \sigma_j(B)}{\abs{\det(B)}} \leq \prod_{j=1}^{n-1} \sigma_j(B) \\
&\leq \norm{B}^{n-1} \leq \norm{B}_F^{n-1} \leq (2^\ell n)^{n-1}.
\end{split}
\end{equation}

Combining the above inequalities, we get
\begin{equation}
\norm{v}_\infty < (2^{Cn}+\tfrac{1}{2}) (2n) 2^{\ell n + n\log_2 n} \leq M',
\end{equation}
and hence 
\begin{equation}
\begin{split}
&\frac{\abs{Q\cap (Q+w)}}{(2M+1)^n} \geq \frac{(2M+1-M')^n}{(2M+1)^n} \\
&\qquad\qquad\geq \Bigl( 1-\frac{1}{(2n)2^\kappa} \Bigr)^n \geq 1-\frac{1}{2^{\kappa+1}}.
\end{split}
\end{equation}
Combining with eq.~(\ref{eqn-overlap-2}), this finishes the proof of eq.~(\ref{eqn-overlap}).


\subsection{Autocorrelation of $f$, and Proof of Eq.~(\ref{eqn-overlap-lb})}
\label{subsec-autocorrelation}

We now prove eq.~(\ref{eqn-overlap-lb}), which shows that the autocorrelation function of $f$ can be approximated by a mixture of Gaussians. First we write 
\begin{equation}
\begin{split}
\bra{f}S_s\ket{f} &= \bra{f}S_s T_\text{scale}^\dagger T_\text{scale} \ket{f} \\
&= \bra{f} T_\text{scale}^\dagger (S_s\tensor I) T_\text{scale} \ket{f} \\
&= \sum_a \abs{q_a}^2 \bra{f^{(a)}}S_s\ket{f^{(a)}}.
\end{split}
\end{equation}
(This follows because the operator $T_\text{scale}$ is block-diagonal, and the operator $S_s$ is diagonal, with respect to the Fourier basis.) 

Next we write 
\begin{equation}
\begin{split}
\abs{q_a}^2 &\bra{f^{(a)}} S_s \ket{f^{(a)}} \\
&= \abs{q_a}^2 \int_{\RR^n} \abs{\widehat{f^{(a)}}(k)}^2 e^{-2\pi ik\cdot s} dk \\
&= \int_{1/(e\lambda a)}^{1/(\lambda a)} \int_{S^{n-1}} F_0(r)^2 W(\lambda ar)^2 \cdot \\
&\qquad \exp(-2\pi i\norm{s}_2 r\cos\phi_1) d\sigma(\phi) r^{n-1} dr \\
&= \int_{1/(e\lambda a)}^{1/(\lambda a)} \int_{S^{n-1}} F_0(r)^2 W(\lambda ar)^2 \cdot \\
&\qquad \cos(2\pi \norm{s}_2 r\cos\phi_1) d\sigma(\phi) r^{n-1} dr,
\end{split}
\end{equation}
using elementary properties of the Fourier transform, writing $k$ in spherical coordinates centered around the direction $s/\norm{s}_2$, and using a change of variables $\phi_1 \mapsto \pi-\phi_1$.

Using the integral representation of the Bessel function $J_\nu(z)$ (formula 9.1.20 in \cite{abramowitz1964handbook}), and letting $S'_0$ be the surface area of the sphere $S^{n-2}$, we get
\begin{equation}
\label{eqn-bessel-1}
\begin{split}
\abs{q_a}^2 &\bra{f^{(a)}} S_s \ket{f^{(a)}} 
= \int_{1/(e\lambda a)}^{1/(\lambda a)} F_0(r)^2 W(\lambda ar)^2 \cdot \\
& J_{(n-2)/2} (2\pi \norm{s}_2 r) \cdot 
\frac{\sqrt{\pi} \Gamma((n-1)/2)}{(\pi \norm{s}_2 r)^{(n-2)/2}} \cdot S'_0 r^{n-1} dr.
\end{split}
\end{equation}

We will use the following approximation: for $\nu$ large (and real), and $0 < x < o(\sqrt{\nu})$, 
\begin{equation}
\label{eqn-bessel-2}
J_\nu(\sqrt{\nu}x) \approx (\sqrt{\nu}x/2)^\nu \frac{1}{\Gamma(\nu+1)} \exp(-x^2/4).
\end{equation}
This is a straightforward consequence of Meissel's formula for $J_\nu(\nu z)$ (\cite{watson1922treatise}, p.227). It can also be derived \cite{stack2018bessel} from Debye's asymptotic expansion for $J_\nu(\nu\,\text{sech}\,\alpha)$ (formula 9.3.7 in \cite{abramowitz1964handbook}). 

Let $\nu = (n-2)/2$ and $x = 2\pi \norm{s}_2 r / \sqrt{(n-2)/2}$. Then eq.~(\ref{eqn-bessel-2}) is valid whenever $0 < \norm{s}_2 r < o(n)$. In addition, we will use the identity 
\begin{equation}
\label{eqn-s0-ratio}
\frac{S'_0}{S_0} = \frac{\Gamma(n/2)}{\sqrt{\pi}\Gamma((n-1)/2)}, 
\end{equation}
where $S_0$ is the surface area of the sphere $S^{n-1}$. Plugging eqs.~(\ref{eqn-bessel-2}) and (\ref{eqn-s0-ratio}) into (\ref{eqn-bessel-1}), we get 
\begin{equation}
\begin{split}
\abs{q_a}^2 &\bra{f^{(a)}} S_s \ket{f^{(a)}} 
\approx \int_{1/(e\lambda a)}^{1/(\lambda a)} F_0(r)^2 W(\lambda ar)^2 \cdot \\
& \exp(-2\pi^2 \norm{s}_2^2 r^2 / (n-2)) \cdot S_0 r^{n-1} dr.
\end{split}
\end{equation}
This approximation is accurate when $\norm{s}_2 < o(n/r)$. When $\norm{s}_2$ is larger than that, the right hand side has an additive error of size $\abs{q_a}^2 \exp(-\Omega(n))$. This proves eq.~(\ref{eqn-overlap-lb}). 


\section{Outlook}
\label{sec-outlook}

The results shown in this paper can be interpreted as evidence that lattice problems are hard for quantum computers. There are a number of directions for further investigation. 

First, these results show that applying the quantum curvelet transform to \textit{single} copies of the state in eq.~(\ref{eqn-index-erasure}) is unlikely to succeed in solving hard lattice problems. However, this still leaves open the possibility that a more sophisticated strategy, involving entangled operations on \textit{multiple} copies of the state in eq.~(\ref{eqn-index-erasure}), could perform better. Such strategies have been studied for other problems, such as the dihedral and symmetric hidden subgroup problems \cite{childs2010quantum}. It is an interesting question whether such a strategy might be able to solve BDD and approximate-SVP --- or whether it too will encounter obstacles that are generalizations of the obstacle described in this paper.

Second, these results work for general lattices. But one might ask whether quantum algorithms can perform better on lattices that have some special structure. For instance, for a class of algebraically-structured lattices known as ideal lattices, there are polynomial-time quantum algorithms that can solve $\gamma$-approximate SVP where $\gamma$ is sub-exponentially large \cite{cramer2017short}. More recently, efficient quantum algorithms have been developed for solving BDD with sub-exponential approximation ratios, for lattices that have abelian group structure \cite{eldar2022efficient}. These algorithms are quite different from our approach using the curvelet transform, and they might not encounter the obstacles studied in this paper.

Finally, it is an interesting question whether our techniques could be applied in the context of other recent work using quantum lattice superposition states, for instance, relating to the continuous LWE problem \cite{bruna2021continuous}, and the superposition LWE problem \cite{chen2022quantum}.


{\vskip 0.3in}

\subsection*{Acknowledgements}

Y.K.L.~thanks Oded Regev and Urmila Mahadev for helpful discussions that motivated this work; and Victor Albert, Gorjan Alagic and Alexander Poremba for helpful comments and suggestions. Y.K.L.~acknowledges support from NIST, DOE ASCR (Fundamental Algorithmic Research for Quantum Computing, award No.~DE-SC0020312), and ARO (Quantum Algorithms for Algebra and Discrete Optimization, grant number W911NF-20-1-0015).


\bibliographystyle{alpha}
\bibliography{cu-bib}

\newcommand{\etalchar}[1]{$^{#1}$}
\begin{thebibliography}{AMRR11}

\bibitem[AMRR11]{ambainis2011symmetry}
Andris Ambainis, Lo{\"\i}ck Magnin, Martin Roetteler, and J{\'e}r{\'e}mie
  Roland.
\newblock Symmetry-assisted adversaries for quantum state generation.
\newblock In {\em 2011 IEEE 26th Annual Conference on Computational
  Complexity}, pages 167--177. IEEE, 2011.

\bibitem[AR03]{aharonov2003lattice}
Dorit Aharonov and Oded Regev.
\newblock A lattice problem in quantum {NP}.
\newblock In {\em 44th Annual IEEE Symposium on Foundations of Computer
  Science, 2003. Proceedings.}, pages 210--219. IEEE, 2003.

\bibitem[AR05]{aharonov2005lattice}
Dorit Aharonov and Oded Regev.
\newblock Lattice problems in {NP} intersect co-{NP}.
\newblock {\em Journal of the ACM (JACM)}, 52(5):749--765, 2005.

\bibitem[AS64]{abramowitz1964handbook}
Milton Abramowitz and Irene~A Stegun.
\newblock {\em Handbook of mathematical functions with formulas, graphs, and
  mathematical tables}, volume~55.
\newblock US Government printing office, 1964.

\bibitem[B{\etalchar{+}}97]{ball1997elementary}
Keith Ball et~al.
\newblock An elementary introduction to modern convex geometry.
\newblock {\em Flavors of geometry}, 31(1--58):26, 1997.

\bibitem[Bab86]{babai1986lovasz}
L{\'a}szl{\'o} Babai.
\newblock On {L}ov{\'a}sz’ lattice reduction and the nearest lattice point
  problem.
\newblock {\em Combinatorica}, 6:1--13, 1986.

\bibitem[BRST21]{bruna2021continuous}
Joan Bruna, Oded Regev, Min~Jae Song, and Yi~Tang.
\newblock Continuous {LWE}.
\newblock In {\em Proceedings of the 53rd Annual ACM SIGACT Symposium on Theory
  of Computing}, pages 694--707, 2021.

\bibitem[CD05]{candes2005continuous}
Emmanuel~J Candes and David~L Donoho.
\newblock Continuous curvelet transform: I. resolution of the wavefront set.
\newblock {\em Applied and Computational Harmonic Analysis}, 19(2):162--197,
  2005.

\bibitem[CDDY06]{candes2006fast}
Emmanuel Candes, Laurent Demanet, David Donoho, and Lexing Ying.
\newblock Fast discrete curvelet transforms.
\newblock {\em multiscale modeling \& simulation}, 5(3):861--899, 2006.

\bibitem[CDW17]{cramer2017short}
Ronald Cramer, L{\'e}o Ducas, and Benjamin Wesolowski.
\newblock Short {S}tickelberger class relations and application to
  {I}deal-{SVP}.
\newblock In {\em Annual International Conference on the Theory and
  Applications of Cryptographic Techniques}, pages 324--348. Springer, 2017.

\bibitem[CLZ22]{chen2022quantum}
Yilei Chen, Qipeng Liu, and Mark Zhandry.
\newblock Quantum algorithms for variants of average-case lattice problems via
  filtering.
\newblock In {\em Annual International Conference on the Theory and
  Applications of Cryptographic Techniques}, pages 372--401. Springer, 2022.

\bibitem[CVD10]{childs2010quantum}
Andrew~M Childs and Wim Van~Dam.
\newblock Quantum algorithms for algebraic problems.
\newblock {\em Reviews of Modern Physics}, 82(1):1, 2010.

\bibitem[EH22]{eldar2022efficient}
Lior Eldar and Sean Hallgren.
\newblock An efficient quantum algorithm for lattice problems achieving
  subexponential approximation factor.
\newblock {\em arXiv preprint arXiv:2201.13450}, 2022.

\bibitem[GR02]{grover2002creating}
Lov Grover and Terry Rudolph.
\newblock Creating superpositions that correspond to efficiently integrable
  probability distributions.
\newblock {\em arXiv preprint quant-ph/0208112}, 2002.

\bibitem[Kho05]{khot2005hardness}
Subhash Khot.
\newblock Hardness of approximating the shortest vector problem in lattices.
\newblock {\em Journal of the ACM (JACM)}, 52(5):789--808, 2005.

\bibitem[Liu09]{liu2009quantum}
Yi-Kai Liu.
\newblock Quantum algorithms using the curvelet transform.
\newblock In {\em Proceedings of the forty-first annual ACM symposium on Theory
  of computing}, pages 391--400, 2009.

\bibitem[Liu10]{liu2010talk}
Yi-Kai Liu.
\newblock Preparing lattice superpositions on a quantum computer.
\newblock Workshop on Post-Quantum Information Security, Joint Quantum
  Institute, University of Maryland, College Park, MD, USA, 2010.

\bibitem[LM09]{lyubashevsky2009bounded}
Vadim Lyubashevsky and Daniele Micciancio.
\newblock On bounded distance decoding, unique shortest vectors, and the
  minimum distance problem.
\newblock In {\em Annual International Cryptology Conference}, pages 577--594.
  Springer, 2009.

\bibitem[MG02]{micciancio2002complexity}
Daniele Micciancio and Shafi Goldwasser.
\newblock {\em Complexity of lattice problems: a cryptographic perspective},
  volume 671.
\newblock Springer Science \& Business Media, 2002.

\bibitem[NC00]{nielsen2000quantum}
Michael~A Nielsen and Isaac~L Chuang.
\newblock {\em Quantum computation and quantum information}.
\newblock Cambridge university press, 2000.

\bibitem[NV10]{nguyen2010lll}
Phong~Q Nguyen and Brigitte Vall{\'e}e.
\newblock {\em The LLL algorithm}.
\newblock Springer, 2010.

\bibitem[P{\etalchar{+}}16]{peikert2016decade}
Chris Peikert et~al.
\newblock A decade of lattice cryptography.
\newblock {\em Foundations and trends{\textregistered} in theoretical computer
  science}, 10(4):283--424, 2016.

\bibitem[Reg09]{regev2009lattices}
Oded Regev.
\newblock On lattices, learning with errors, random linear codes, and
  cryptography.
\newblock {\em Journal of the ACM (JACM)}, 56(6):1--40, 2009.

\bibitem[sta18]{stack2018bessel}
Limit of a {B}essel function is a {G}aussian.
\newblock Mathematics Stack Exchange,
  \url{https://math.stackexchange.com/questions/2974428/limit-of-a-bessel-function-is-a-gaussian},
  2018.

\bibitem[Wat22]{watson1922treatise}
George~Neville Watson.
\newblock {\em A treatise on the theory of Bessel functions}, volume~2.
\newblock The University Press, 1922.

\end{thebibliography}


\appendix
\section{Upper Bounds on the Uncertainty of the Curvelet Transform}
\label{sec-finding-center-upper-bounds}

We now prove eq.~(\ref{eqn-curvelet-upper-bound}), which upper-bounds the uncertainty of the measurement outcomes that are obtained from the wavefunction returned by the quantum curvelet transform.

More precisely, we show that there exists a family of curvelets $\gamma_{a,b,\theta}$, and there exists a constant $C'>0$, such that for any radial function $f$ centered at $c \in \RR^n$, and for any $0<a\leq 1$, 
\begin{equation}
\label{eqn-curvelet-upper-bound-int}
\begin{split}
\int_{S^{n-1}} &\int_{\RR^n} (b-c)^T(I-\theta\theta^T)(b-c) \abs{\Gamma_f(a,b,\theta)}^2 db d\sigma(\theta) \\
&\leq \tfrac{1}{\sqrt{2}}a \int_{\RR^n} \norm{x-c}_2^2 \abs{\Xi_f(a,x)}^2 dx \\
&+ C'\lambda^2 n^2 a \int_{S^{n-1}} \int_{\RR^n} \abs{\Gamma_f(a,b,\theta)}^2 db d\sigma(\theta).
\end{split}
\end{equation}
Note that eq.~(\ref{eqn-curvelet-upper-bound-int}) implies eq.~(\ref{eqn-curvelet-upper-bound}), by computing conditional probabilities, and making use of eq.~(\ref{eqn-xi-f-plancherel-a}). 

We now prove eq.~(\ref{eqn-curvelet-upper-bound-int}). Without loss of generality, let $c = \vec{0} \in \RR^n$. We proceed as in Section \ref{sec-simplifying-integrals}: let $T$ denote the left-hand side of eq.~(\ref{eqn-uncert-1}), which is the same as the left-hand side of eq.~(\ref{eqn-curvelet-upper-bound-int}). Then write 
\begin{equation}
\label{eqn-T-copy}
T = \frac{S_0 (n-1)}{(2\pi)^2} (I_A + 2I_B + I_C),
\end{equation}
using eq.~(\ref{eqn-T}), and the accompanying definitions of $I_A$, $I_B$ and $I_C$. We will upper-bound each of these terms. 

For $I_B$ and $I_A$, we proceed in a similar manner as in Sections \ref{sec-lower-bounding-IB} and \ref{sec-lower-bounding-IA}. For $I_C$, the calculation is more involved, and depends on the choice of the angular window function $V$ in the definition of the curvelet transform.


\subsection{Upper-Bounding $I_B$}

As in Section \ref{sec-lower-bounding-IB}, eqs.~(\ref{eqn-IB}) and (\ref{eqn-I2-calc}), we write
\begin{equation}
\label{eqn-IB-copy}
I_B = I_{Br} I_{B1} I_2,
\end{equation}
where $I_2$ is given by
\begin{equation}
I_2 = \frac{S'_0}{n-1}, \quad S'_0 = \int_{S^{n-2}} d\sigma(\phi_2,\ldots,\phi_{n-1}).
\end{equation}

Recall the upper-bound on $I_{B1}$ in eq.~(\ref{eqn-IB1-calc}). A similar argument yields a lower-bound on $I_{B1}$ (which matches the upper-bound up to a constant factor):
\begin{equation}
I_{B1} \geq -\frac{n-1}{2S'_0}.
\end{equation}

Recall the upper-bound on $I_{Br}$ in eq.~(\ref{eqn-IBr-calc}). A similar argument yields a lower-bound on $I_{Br}$ (which matches the upper-bound up to a constant factor):
\begin{equation}
\label{eqn-IBr-calc-2}
\begin{split}
I_{Br} &= -\tfrac{1}{2} (n-2) \int_0^\infty H_0(r)^2 r^{n-3} dr \\
&= -\tfrac{1}{2} (n-2) \int_0^\infty F_0(r)^2 W(\lambda ar)^2 r^{n-3} dr \\
&\geq -\tfrac{1}{2} (n-2) e^2\lambda^2 a^2 \int_0^\infty F_0(r)^2 W(\lambda ar)^2 r^{n-1} dr.
\end{split}
\end{equation}

Substituting into eq.~(\ref{eqn-IB-copy}), we get
\begin{equation}
I_B \leq \tfrac{1}{4} (n-2) e^2\lambda^2 a^2 \int_0^\infty F_0(r)^2 W(\lambda ar)^2 r^{n-1} dr.
\end{equation}
Hence we have:
\begin{multline}
\tfrac{S_0 (n-1)}{2\pi^2} I_B \leq \tfrac{1}{8\pi^2} S_0 (n-1) (n-2) e^2\lambda^2 a^2 \cdot \\
\int_0^\infty F_0(r)^2 W(\lambda ar)^2 r^{n-1} dr.
\end{multline}
Then using eq.~(\ref{eqn-curvelet-plancherel-a}), we get
\begin{multline}
\tfrac{S_0 (n-1)}{2\pi^2} I_B \leq \tfrac{1}{8\pi^2} (n-1) (n-2) e^2\lambda^2 a^2 \cdot \\
\int_{S^{n-1}} \int_{\RR^n} \abs{\Gamma_f(a,b,\theta)}^2 db d\sigma(\theta).
\end{multline}
We substitute this into eq.~(\ref{eqn-T-copy}), to upper-bound $T$.


\subsection{Upper-Bounding $I_A$}

As in Section \ref{sec-lower-bounding-IA}, eq.~(\ref{eqn-IA}), we write
\begin{equation}
\label{eqn-IA-copy}
I_A = I_{Ar} I_{A1} I_2.
\end{equation}
As before, we have $I_2 = \frac{S'_0}{n-1}$.

To upper-bound $I_{A1}$, recall that the angular window function $V$ is supported on $[0, \tfrac{1}{\sqrt{2}}]$, so we have 
\begin{equation}
\begin{split}
I_{A1} &= \int_0^{\sin^{-1}(\sqrt{a/2})} V(\tfrac{1}{\sqrt{a}} \sin\phi_1)^2 
C_{a,n}^2 \sin^n\phi_1 d\phi_1.
\end{split}
\end{equation}
Inside the integral, we have $\cos\phi_1 \geq \frac{1}{\sqrt{2}}$, and $\sin\phi_1 \geq 0$, hence 
\begin{multline}
I_{A1} \leq \sqrt{2} \int_0^{\sin^{-1}(\sqrt{a/2})} V(\tfrac{1}{\sqrt{a}} \sin\phi_1)^2 \cdot \\
C_{a,n}^2 \cos\phi_1 \sin^n\phi_1 d\phi_1.
\end{multline}
After changing variables (letting $y = \tfrac{1}{\sqrt{a}} \sin\phi_1$), using eq.~(\ref{eqn-normalization-bounds}) to bound $C_{a,n}^2$, and using the definition of $M_{n-2}$ in eq.~(\ref{eqn-M-n-2}), we get
\begin{equation}
\begin{split}
I_{A1}
&\leq \sqrt{2} C_{a,n}^2 a^{(n+1)/2} \int_0^{1/\sqrt{2}} V(y)^2 y^n dy \\
&\leq \frac{1}{\sqrt{2}} C_{a,n}^2 a^{(n+1)/2} \int_0^{1/\sqrt{2}} V(y)^2 y^{n-2} dy \\
&\leq \frac{a}{\sqrt{2}S'_0}.
\end{split}
\end{equation}

Finally, recall the expression for $I_{Ar}$ in eq.~(\ref{eqn-IAr-calc}):
\begin{equation}
I_{Ar} = \frac{(2\pi)^2}{S_0} \int_{\RR^n} \norm{x}_2^2 \abs{\Xi_f(a,x)}^2 dx.
\end{equation}

Substituting into eq.~(\ref{eqn-IA-copy}), we get
\begin{equation}
I_A \leq 
\frac{(2\pi)^2}{S_0} \frac{a}{\sqrt{2}(n-1)} \int_{\RR^n} \norm{x}_2^2 \abs{\Xi_f(a,x)}^2 dx.
\end{equation}
Hence we have:
\begin{equation}
\frac{S_0 (n-1)}{(2\pi)^2} I_A \leq
\frac{a}{\sqrt{2}} \int_{\RR^n} \norm{x}_2^2 \abs{\Xi_f(a,x)}^2 dx.
\end{equation}
We substitute this into eq.~(\ref{eqn-T-copy}), to upper-bound $T$.


\subsection{Upper-Bounding $I_C$}

We can write $I_C$ as a product of three factors:
\begin{equation}
\label{eqn-IC}
I_C = I_{Cr} I_{C1} I_2,
\end{equation}
\begin{equation}
I_{Cr} = \int_0^\infty H_0(r)^2 r^{n-3} dr,
\end{equation}
\begin{equation}
I_{C1} = \int_0^\pi X'_0(\phi_1)^2 \cos^2\phi_1 \sin^{n-2}\phi_1 d\phi_1,
\end{equation}
where $I_2 = \frac{S'_0}{n-1}$.

To upper-bound $I_{C1}$, recall that the angular window function $V$ is supported on $[0, \tfrac{1}{\sqrt{2}}]$, so we have 
\begin{multline}
I_{C1} = \int_0^{\sin^{-1}(\sqrt{a/2})} V'(\tfrac{1}{\sqrt{a}} \sin\phi_1)^2 \cdot \\
\frac{C_{a,n}^2}{a} \cos^4\phi_1 \sin^{n-2}\phi_1 d\phi_1.
\end{multline}
Inside the integral, we have $1 \geq \cos\phi_1 \geq \frac{1}{\sqrt{2}}$ and $\sin\phi_1 \geq 0$, hence 
\begin{multline}
I_{C1} \leq \int_0^{\sin^{-1}(\sqrt{a/2})} V'(\tfrac{1}{\sqrt{a}} \sin\phi_1)^2 \cdot \\
\frac{C_{a,n}^2}{a} \cos\phi_1 \sin^{n-2}\phi_1 d\phi_1.
\end{multline}
After changing variables (letting $y = \tfrac{1}{\sqrt{a}} \sin\phi_1$), using eq.~(\ref{eqn-normalization-bounds}) to bound $C_{a,n}^2$, using the definition of $M_{n-2}$ in eq.~(\ref{eqn-M-n-2}), and defining 
\begin{equation}
\label{eqn-Mprime-n-2}
M'_{n-2} = \int_0^{1/\sqrt{2}} V'(y)^2 y^{n-2} dy,
\end{equation}
we get
\begin{equation}
\begin{split}
I_{C1}
&\leq C_{a,n}^2 a^{(n-3)/2} \int_0^{1/\sqrt{2}} V'(y)^2 y^{n-2} dy \\
&\leq \frac{M'_{n-2}}{M_{n-2}} \, \frac{1}{a S'_0}.
\end{split}
\end{equation}

We can upper-bound $I_{Cr}$ in the same way that we lower-bounded $I_{Br}$ in eq.~(\ref{eqn-IBr-calc-2}):
\begin{equation}
\begin{split}
I_{Cr} &= \int_0^\infty H_0(r)^2 r^{n-3} dr \\
&= \int_0^\infty F_0(r)^2 W(\lambda ar)^2 r^{n-3} dr \\
&\leq e^2\lambda^2 a^2 \int_0^\infty F_0(r)^2 W(\lambda ar)^2 r^{n-1} dr.
\end{split}
\end{equation}

Substituting into eq.~(\ref{eqn-IC}), we get
\begin{equation}
I_C \leq \frac{M'_{n-2}}{M_{n-2}} \, \frac{e^2\lambda^2 a}{n-1} \int_0^\infty F_0(r)^2 W(\lambda ar)^2 r^{n-1} dr.
\end{equation}
Hence we have:
\begin{multline}
\frac{S_0 (n-1)}{(2\pi)^2} I_C 
\leq \frac{S_0}{(2\pi)^2} \frac{M'_{n-2}}{M_{n-2}} \, e^2\lambda^2 a \cdot \\
\int_0^\infty F_0(r)^2 W(\lambda ar)^2 r^{n-1} dr.
\end{multline}
Then using eq.~(\ref{eqn-curvelet-plancherel-a}), we get
\begin{multline}
\label{eqn-IC-calc-2}
\frac{S_0 (n-1)}{(2\pi)^2} I_C 
\leq \frac{1}{(2\pi)^2} \frac{M'_{n-2}}{M_{n-2}} \, e^2\lambda^2 a \cdot \\
\int_{S^{n-1}} \int_{\RR^n} \abs{\Gamma_f(a,b,\theta)}^2 db d\sigma(\theta).
\end{multline}

In the above bound, the quantity $M'_{n-2}/M_{n-2}$ depends on the choice of the angular window funciton $V$, in the definition of the curvelet transform. In the following section, we show a simple choice of $V$ for which
\begin{equation}
\label{eqn-M-prime-M-ratio}
\frac{M'_{n-2}}{M_{n-2}} \lesssim \frac{2n^2}{3}.
\end{equation}
We will then substitute eqs.~(\ref{eqn-IC-calc-2}) and (\ref{eqn-M-prime-M-ratio}) into eq.~(\ref{eqn-T-copy}), to upper-bound $T$.


\subsection{Choosing the Angular Window Function $V$}

We describe a simple choice of $V$, such that eq.~(\ref{eqn-M-prime-M-ratio}) holds. First, we define $q$ to be a piecewise quadratic function that increases monotonically from $q(0) = 0$ to $q(2) = 1$:
\begin{equation}
q(x) = \begin{cases}
\tfrac{1}{2}x^2, &0\leq x\leq 1,\\
1 - \tfrac{1}{2}(2-x)^2, &1<x\leq 2.
\end{cases}
\end{equation}
The behavior of $q$ can be understood by looking at its derivative $q'$, which increases monotonically from $q'(0) = 0$ to $q'(1) = 1$, and then decreases monotonically to $q'(2) = 0$:
\begin{equation}
q'(x) = \begin{cases}
x, &0\leq x\leq 1,\\
2-x, &1<x\leq 2.
\end{cases}
\end{equation}

We then define $V$ as follows. For notational convenience, we let $\alpha = 1/\sqrt{2}$. We then define $V$ to be the composition of $q$ with a suitable affine transformation, so that $V$ decreases monotonically from $V(0) = 1$ to $V(\alpha) = 0$:
\begin{equation}
V(x) = q(\tfrac{2}{\alpha}(\alpha-x)), \quad 0\leq x\leq \alpha.
\end{equation}

The quantities $M_{n-2}$ and $M'_{n-2}$ can then be evaluated, via a tedious but elementary calculation, to obtain:
\begin{equation}
M_{n-2} \gtrsim \frac{96}{n^5} \alpha^{n-1},
\end{equation}
\begin{equation}
M'_{n-2} \sim \frac{32}{n^3} \alpha^{n-3}.
\end{equation}
This implies eq.~(\ref{eqn-M-prime-M-ratio}).


\end{document}